\documentclass[aps,prd,onecolumn,groupedaddress,showpacs,nofootinbib,amssymb]{revtex4}
\usepackage[dvips]{graphicx}
\usepackage{amssymb}
\usepackage{amsmath}
\usepackage{graphicx,,color}
\usepackage{amsfonts}
\usepackage{bm}
\usepackage{xcolor}
\usepackage{cancel}
\usepackage{comment}
\newcommand\be{\begin{equation}}
\newcommand\ee{\end{equation}}

\DeclareUnicodeCharacter{2212}{-}

\DeclareMathOperator{\ev}{eV}
\DeclareMathOperator{\erf}{erf}
\allowdisplaybreaks[4]

\begin{document}
\tolerance=5000

\title{Swampland criteria for rescaled Einstein-Hilbert gravity with string corrections}
\author{
Achilles Gitsis,\,\thanks{agitsis62@gmail.com}
Konstantinos-Rafail Revis,\,\thanks{reviskostis@gmail.com}
S.A.
Venikoudis,\,\thanks{venikoudis@gmail.com}
F.P.
Fronimos\, \thanks{fotisfronimos@gmail.com}
}
\affiliation{
 Department of Physics, Aristotle University of
Thessaloniki, Thessaloniki 54124,
Greece\\}

\tolerance=5000

\begin{abstract}
In this work we examine the Swampland criteria for a specific class of rescaled $f(R)$ gravitational models, that are capable of unifying the primordial era of the Universe with the late-time era with the inclusion of string corrections. In particular, we develop separately the theoretical framework of Gauss-Bonnet and Chern-Simons theories considering that, the rescale parameter is constrained to reside in the area $0<\alpha<1$. As showcased, in the context of the aforementioned theories, a viable inflationary phenomenology consistent with the latest Planck data can be obtained
for both cases for a wide variety of values. The Swampland criteria which where examined are satisfied, not necessarily simultaneously, for small values of the rescale parameter, which is in agreement with the case of a canonical scalar field with absent string corrective terms. The Gauss-Bonnet model is also further constrained, in order to obtain a propagation velocity of tensor perturbations which coincides with that of light, according to the recent observations from the GW170817. As a result of this assumption the degrees of freedom of the theory are reduced. An interesting feature which arises from the overall phenomenology is that, due to the inclusion of string corrections the tensor spectral index of primordial perturbations is now capable of obtaining a positive value which is not possible in the case of the canonical scalar field. Last but not least, the power-law model which is known to be incompatible with observations is now rendered viable by including a parity violating term and as showcased, it satisfies the Swampland criteria as well.  
\end{abstract}

\maketitle

\section{Introduction}
String theory is one of the most prominent theories, posing as a strong candidate (theoretically, at least) for the unification of fundamental forces from a particle point of view. However, up to the writing date there is no experimental evidence providing any proof that string theory is a valid theory. This is the reason why many researchers turn their interest to the study of other theories, that may serve as a means in order to give prominence to string theory and a consistent M-theory at a later stage.

One of such theories is the inflationary paradigm, which was firstly developed by Alan Guth \cite{Guth:1980zm}, since it is well believed that the accelerated expansion era, occurred right after the quantum gravity era, when strings were significant. The theoretical framework of  inflation provides solutions to major discrepancies arising from the Big Bang Cosmological Model with the most important of them being, the horizon and the flatness problem of the Universe, the absence of magnetic monopoles and the last but not least the problem of initial entropy. In the present work, we investigate the effect of the constraints, that are imposed by the timely Swampland Criteria on a rescaled Einstein-Hilbert gravitational action. This is practically an extension of one of our previous studies \cite{Oikonomou:2021zfl} with a power-law-like class of modified gravity theories and the inclusion of string-inspired corrective terms. The justification for the use of modified gravitational theories \cite{Nojiri:2017ncd,Capozziello:2011et,Capozziello:2010zz,Nojiri:2006ri,Nojiri:2010wj,Olmo:2011uz} lies in the fact that the inflationary era corresponds to the large curvature regime thus, explaining the need of higher-order terms in the effective Lagrangian \cite{Cheung:2007st}. We focus on canonical scalar-field inflation.
This is the link towards string theory, since scalar fields constitute a representative feature of the aforementioned theory. The main advantage of the Swampland Criteria is that they distinguish the effective field theories that provide us with a consistent M-theory.

Given that, the effective Lagrangian of inflation has not been constrained yet from observations, it is reasonable to assume that higher curvature  corrections such as the Gauss-Bonnet \cite{Kanti:2015pda,Yi:2018gse,Guo:2010jr,Jiang:2013gza,Guo:2009uk,DeLaurentis:2015fea,Fomin:2020hfh,Pozdeeva:2020apf,vandeBruck:2016xvt,Odintsov:2018zhw,Nozari:2017rta,Chakraborty:2018scm,Kawai:1999pw,vandeBruck:2017voa,Bakopoulos:2020dfg,Kleihaus:2019rbg,Bakopoulos:2019tvc,Kanti:1995vq,Bajardi:2019zzs,Venikoudis:2022xyq} may be included, representing the imprint of the quantum era. It is well-known in literature that the Einstein-Gauss-Bonnet theories with the involvement of scalar fields can provide a viable description of the inflationary epoch. However, the Gauss-Bonnet theories predicted
deviated velocity of primordial tensors perturbations with that of light. The solution to this issue was given by the observations of the merging of two neutron stars. Specifically, it was observed that the GW170817, which was produced from this merging had the velocity of the light or almost equal to this. For previous works concerning the compatibility of Einstein-Gauss-Bonnet theories with the GW170817 see Refs. \cite{Odintsov:2020sqy,Venikoudis:2021irr,Venikoudis:2021oee,Odintsov:2020xji}. The main feature of this compatibility is that primordial gravitons must be retained  massless. Even though the GW170817 took place during the late-time cosmological era, from the particle physics point of view, there is no mechanism to explain changes in the graviton's mass in between the primordial and the late-time era of the Universe. In addition, the hypothesis of massless gravitons is essential feature on the novel 4-D Einstein-Gauss-Bonnet gravity, which was firstly proposed by Glavan and Lin, see Ref.\cite{Glavan:2019inb}. According to this scenario the authors rescale the Gauss-Bonnet coupling from $\alpha$ to $\alpha/D-4$ in order to investigate the local dynamics of the theory. Essentially, this kind of dimensional regularization procedure can be used for the extraction of the local dynamics even for modified gravitational theories based on the Ref.\cite{Ai:2020peo}.

Furthermore, the addition of the Chern-Simons parity violating term \cite{Nojiri:2020pqr,Nojiri:2019nar,Odintsov:2019evb,Odintsov:2019mlf,Alexander:2009tp,Qiao:2019hkz,Nishizawa:2018srh,Wagle:2018tyk,Yagi:2012vf,Yagi:2012ya,Molina:2010fb,Izaurieta:2009hz,Smith:2007jm,Konno:2009kg,Sopuerta:2009iy,Matschull:1999he,Haghani:2017yjk,Fronimos:2021czc} in the gravitational action, even thought is absent from the equations of motion, has a strong influence on the tensor perturbations. In this context, a slightly blue-titled tensor index of primordial perturbations can be manifested \cite{Camerini:2008mj}, which is a feature that may be observed the following years from the LISA detector or other interferometer experiments by opening new frontier for theoretical Cosmology. In addition, recently has been proved that the generated spectrum of primordial tensor perturbations is strongly chiral due to the involvement of the Chern-Simons term in the gravitational action, see Ref. \cite{Odintsov:2022hxu}. Overall, a theory with the inclusion of the aforementioned gravitational terms predict primordial gravitational waves with the velocity of the light's with two circular polarization states. 

The paper is constructed as follows. In Sect.II the Swampland criteria are reviewed and the $f(R)$ gravity which shall be assumed is specified. In Sect.III the inflationary phenomenology of the rescaled $f(R)$ gravity in the case of a canonical scalar field non minimally coupled to gravity is presented and the viability of certain models along with the Swampland criteria is examined. Afterwards, in Sect.IV the inflationary dynamics of the same rescaled $f(R)$ model is tested in the case of a Chern-Simons term introduced yet again via a non minimal coupling with the scalar field. Finally in Sect.V the concluding remarks are presented. 

\section{Swampland Criteria for rescaled gravity}
Before we proceed with the study of the string corrected $f(R)$ gravity, we shall briefly cover the Swampland criteria. Swampland conjectures where firstly introduced in Refs. \cite{Vafa:2005ui,Ooguri:2006in} and they have been extensively studied in Refs. \cite{Palti:2020qlc,Mizuno:2019bxy,Brandenberger:2020oav,Blumenhagen:2019vgj,Wang:2019eym,Benetti:2019smr,Palti:2019pca,Cai:2018ebs,Akrami:2018ylq,Mizuno:2019pcm,Aragam:2019khr,Brahma:2019mdd,Mukhopadhyay:2019cai,Yi:2018dhl,Gashti:2022hey,Brahma:2019kch,Haque:2019prw,Heckman:2019dsj,Acharya:2018deu,Elizalde:2018dvw,Cheong:2018udx,Heckman:2018mxl,Kinney:2018nny,Garg:2018reu,Lin:2018rnx,Park:2018fuj,Olguin-Tejo:2018pfq,Fukuda:2018haz,Wang:2018kly,Ooguri:2018wrx,Matsui:2018xwa,Obied:2018sgi,Agrawal:2018own,Murayama:2018lie,Marsh:2018kub,Storm:2020gtv,Trivedi:2020wxf,Sharma:2020wba,Odintsov:2020zkl,Mohammadi:2020twg,Trivedi:2020xlh}. The aforementioned criteria serve as conditions, that suggest whether an effective field theory is not a proper description for quantum gravity at high energy scales. It should also be stated that, unlike Ref. \cite{Benetti:2019smr} where the criteria for the $f(R)$ gravity were derived by connecting the Jordan and Einstein frame via a conformal transformation, in the present article we shall assume that the $f(R)$ gravity part of the gravitational action becomes dominant in the late-time era and thus the criteria that will be discussed apply to the scalar field and its canonical potential and not the form of the $f(R)$ gravity. Now, an effective theory that respects the Weak Gravity Conjecture must  satisfy the following criterions:
\begin{itemize}
  \item The Swampland distance conjecture. It suggests that the effective theory has a specific field range,
  \begin{equation}
  \centering
  \label{criterion1}
  |\kappa\Delta\phi|<\mathcal{O}(1)\, ,
  \end{equation}
  \item The de Sitter conjecture. According to this condition and for a positively valued canonical scalar potential, its derivative must  have a lower bound,
  \begin{equation}
  \centering
  \label{criterion2}
  \frac{|V'(\phi_i)|}{\kappa V(\phi_i)}>\mathcal{O}(1)
  \end{equation}
   An equivalent condition that the scalar potential may satisfy reads,
  \begin{equation}
  \centering
  \label{criterion3}
  -\frac{V''(\phi_i)}{\kappa^2V(\phi_i)}>\mathcal{O}(1)\, ,
  \end{equation}
\end{itemize}
where the prime denotes differentiation with respect to the scalar field $\phi$, $\kappa=\frac{1}{M_P}$ with $M_P$ being the reduced Planck mass and $\phi_i$ is the value of the scalar field during the first horizon crossing. In the following models we shall study the case of $\frac{V'(\phi_i)}{\kappa V(\phi_i)}>\mathcal{O}(1)$. These are the conditions that are satisfied for an effective theory that is not UV complete. It should be stated that the conditions are indicative of the UV completeness and do not need to be respected simultaneously in order to obtain a UV incomplete effective theory.

Let us now proceed with the appropriate form of the $f(R)$ gravity. All the models that are examined in the present article belong to the general category of the form,
\begin{equation}
\label{fR}
\centering
f(R)=R-c_1Re^{-\frac{R_0}{R}}-R_1\bigg(\frac{R}{R_2}\bigg)^n-c_2\bigg(\frac{R}{M_P}\bigg)^2\ln\bigg(1+\frac{R_3 }{R}\bigg)-R_4
\end{equation}
with $c_i's$ being dimensionless parameters and the $R_i^{'}s$ being arbitrary parameters with mass dimensions $\ev^2$. In addition, R stands for the Ricci scalar and the auxiliary exponent satisfies the condition $n<1$. The form of the $f(R)$ indicates that it becomes dominant during the late-time era where $R\xrightarrow{}0$ however, during the inflationary era the prefactor of the linear Ricci scalar is shifted as suggested by the Taylor expansion of the exponential and the logarithm in the limit of $R \xrightarrow{} \infty$.

In particular, the dominant contribution in the early era now reads $f(R)\simeq\alpha R$ where $\alpha=1-\left(c_1+c_2\frac{R_3}{M_P^2}\right)$ is the rescale parameter where it is assumed that $R_3<M_P^2$ such that $\alpha$ is non negative. In principle, apart from the exponential and logarithmic function, other functions may be considered, for instance $c_1 R\cosh\bigg(\frac{R_0}{R}\bigg)$ affects $\alpha$ in the same manner as the exponential or $\bigg(\frac{R}{M_P}\bigg)^2\sinh\bigg(\frac{R_3}{R}\bigg)$ and $\bigg(\frac{R}{M_P}\bigg)^2\erf\bigg(\frac{R_3}{R}\bigg)$ affect such parameter exactly as the logarithm.
Other hyperbolic or special functions are also a viable option.

Before we proceed any further it should be stated that throughout this paper we shall assume a flat and homogeneous background which corresponds to the Friedmann-Robertson-Walker line element,
\begin{equation}
\centering
\label{metric}
ds^2=-dt^2+a^2(t)\delta_{ij}dx^idx^j,\,
\end{equation}
where $a(t)$ stands for the scale factor. In consequence, it is reasonable to assume that the scalar field is solely time dependent.

\section{Theoretical Framework of Rescaled $f(R)$ Einstein-Gauss-Bonnet gravity}
Our analysis commences from the gravitational action, which corresponds to a rescaled $f(R)$ model in the presence of a canonical scalar field and string corrective terms. In particular, string-inspired corrections are expressed through a non minimal coupling between the scalar field and higher order curvature terms namely, the Gauss-Bonnet topological invariant. According to this assumption, the resulting gravitational action obtains the following form,
\begin{equation}
\centering
\label{actionGB}
\mathcal{S}=\int{d^4x\sqrt{-g}\left(\frac{\alpha R}{2\kappa^2}-\frac{1}{2}g^{\mu\nu}\nabla_\mu\phi\nabla_\nu\phi-V(\phi)-\xi(\phi)\mathcal{G}\right)}\, ,
\end{equation}
where $\alpha$ is the dimensionless parameter specified above, which takes values in the range $0<\alpha<1$. In addition, g is the determinant of the metric tensor and
$\frac{1}{2}g^{\mu\nu}\nabla_\mu\phi\nabla_\nu\phi$, $V(\phi)$ stand for the kinetic term and the potential of the scalar field respectively. The higher order curvature correction, denoted as $\mathcal{G}$, is defined as $\mathcal{G}=R^2-4R_{\alpha\beta}R^{\alpha\beta}+R_{\alpha\beta\gamma\delta}R^{\alpha\beta\gamma\delta}$,
with $R_{\alpha\beta}$ and $R_{\alpha\beta\gamma\delta}$ being the
Ricci and Riemann tensor respectively. Finally, $\xi(\phi)$ is an arbitrary, for the time being, dimensionless scalar coupling function assuming that it satisfies a particular differential equation for reasons that will become clear subsequently. In order to extract the equations of the motion of the theory, we implement the variational principle in the gravitational action (\ref{actionGB}) with respect to the metric tensor and the scalar field. As a result, the equations of motion read,
\begin{equation}
\centering
\label{motion1GB}
\frac{3\alpha H^2}{\kappa^2}=\frac{1}{2}\dot\phi^2+V+24\dot\xi H^3,\,
\end{equation}
\begin{equation}
\centering
\label{motion2GB}
-\frac{2 \alpha \dot H}{\kappa^2}=\dot\phi^2-16\dot\xi H\dot H-8H^2(\ddot\xi-H\dot\xi),\,
\end{equation}
\begin{equation}
\centering
\label{motion3GB}
\ddot\phi+3H\dot\phi+V'+\xi'\mathcal{G}=0.\
\end{equation}
It is interesting to be mentioned that the rescale parameter $\alpha$, although it sees to be absent from the continuity equation of the scalar field (\ref{motion3GB}), it essentially affects its evolution with respect to time, since it participates in the Friedmann's and Raychadhuri's equations (\ref{motion1GB}-\ref{motion2GB}). This suggests that all observational indices that we shall study in the present article are affected by such parameter.

Before we proceed with the overall formalism, let us briefly discuss the impact of the Gauss-Bonnet scalar coupling function $\xi(\phi)$. Initially, it is used in order to examine the contribution from higher order curvature terms, meaning the topological invariant $\mathcal{G}$, through a non minimal coupling with the scalar field, since it would vanish in 4D as a total derivative. However, such contribution is known for affecting tensor perturbations and in particular their propagation velocity, as it is described in Ref. \cite{Hwang:2005hb}. In other words, primordial gravitational waves do not necessarily propagate with the velocity of light, assuming that $\xi(\phi)$ is indeed dynamical and not constant. The aforementioned velocity is given by the following expression,
\begin{equation}
\centering
\label{cTeq}
c_\mathcal{T}^2=1-\frac{Q_f}{Q_{GB}},\,
\end{equation}
where $Q_f= 8(\ddot\xi-H\dot\xi)$ and $Q_{GB}=\frac{\alpha}{\kappa^2}-8\dot \xi H$ are auxiliary functions. The velocity of gravitational waves is given in Natural units and as indicated, it deviates from the velocity of light by a factor of $\frac{Qf}{Q_{GB}}$. In this context, the rescaled $f(R)$ Einstein-Gauss-Bonnet model would assume a non trivial propagation velocity of tensor perturbations and as a result a massive primordial graviton would be expected in this low-energy effective model. In order to avoid this, the propagation velocity $c_T$ is constrained and becomes equal to unity by imposing the condition $\ddot \xi=H \dot \xi$, as shown in Ref. \cite{Odintsov:2020sqy}, by indicating compatibility with the GW170817 event, for further details see Ref. \cite{LIGOScientific:2017vwq,LIGOScientific:2017ync,LIGOScientific:2017zic,LIGOScientific:2018cki,LIGOScientific:2018hze,Ezquiaga:2017ekz,Sakstein:2017xjx,LIGOScientific:2018dkp}. In turn, this condition serves now as a differential equation for the Gauss-Bonnet scalar coupling function which effectively decreases the total number of degrees of freedom of the theory. To indicate this properly, let us implement the slow-roll conditions, 
\begin{align}
\label{approx}
\centering
\dot H&\ll H^2,& \frac{1}{2}\dot\phi^2&\ll V,& \ddot\phi\ll3 H\dot\phi,\
\end{align}
into the differential equation $\ddot \xi=H \dot \xi$. According to the chain rule, the first and the second time derivatives of the Gauss-Bonnet scalar coupling function are $\dot \xi=\dot \phi \xi'$ and $\ddot \xi=\ddot \phi \xi'+\xi'' \dot \phi^2$ respectively  
where, the prime stands for differentiation with respect to the scalar field $\xi'=\frac{d\xi}{d\phi}$.
Substituting both time derivatives of the scalar coupling function into the differential equation, we get that,
\begin{equation}
\centering
\label{difeq}
\ddot \phi \xi'+\xi'' \dot \phi^2=H\dot \phi \xi' ,
\end{equation}
and finally one can easily ascertain that the time derivative of the homogeneous scalar field has the following expression,
\begin{equation}
\label{fdot}
\centering
\dot \phi \simeq \frac{H\xi'}{\xi''}.
\end{equation}
In the expression above the factor $\frac{\ddot\phi}{H\dot\phi}$ was neglected as a subleading contribution, given that the slow-roll indices during the first horizon crossing are approximately of order $\mathcal{O}(10^{-3})$. There is also the option of keeping such contribution under the constant roll assumption $\ddot\phi=\beta H\dot\phi$, for details see Ref.\cite{Odintsov:2020mkz}, however the constant roll condition shall not be investigated in the present article. Consequently, by imposing the constraint for the speed of the gravitational waves and the slow-roll conditions, the system of equations of motion is reduced into a simpler form,
\begin{equation}
\centering
\label{motion1GBsimple}
\frac{3\alpha H^2}{\kappa^2}\simeq V+24\dot\xi H^3,\,
\end{equation}
\begin{equation}
\centering
\label{motion2GBsimple}
-\frac{2 \alpha \dot H}{\kappa^2}\simeq\dot\phi^2-16\dot\xi H\dot H,\,
\end{equation}
\begin{equation}
\centering
\label{motion3GBsimple}
3H\dot\phi+V'+24\xi' H^4\simeq0.\
\end{equation}
However, even though the imposed approximations holding true, the system of equations of motion remains quite complicated to be solved with an analytic way. For this reason, further approximations are necessary in order to extract information about the inflationary phenomenology. More specifically, we consider that the string corrections, which are involved in the equations of motion, have negligible numerical contribution thus, we neglect them. The contribution however of the Gauss-Bonnet scalar coupling function remains in the time derivative of the scalar field as shown in Eq.(\ref{fdot}). Therefore, the simplified system now obtains the following form,
\begin{equation}
\centering
\label{motion1GBfinal}
H^2\simeq \frac{\kappa^2 V}{3\alpha},\,
\end{equation}
\begin{equation}
\centering
\label{motion2GBfinal}
\dot H\simeq - \frac{H^2}{2\alpha}\bigg(\frac{\kappa \xi'}{\xi''}\bigg)^2,\,
\end{equation}
\begin{equation}
\centering
\label{motion3GBfinal}
V'+\frac{\xi'}{\xi''}\frac{\kappa^2 V}{\alpha}+\frac{8}{3\alpha^2}\xi'\kappa^4V^2\simeq 0.\
\end{equation}
According to the above expressions, it becomes abundantly clear that the rescale parameter $\alpha$ indeed affects the evolution of the scalar field as it was suggested before.
In order to investigate the dynamics of the early Universe, we introduce the slow-roll indices based on Ref. \cite{Hwang:2005hb},
\begin{align}
\centering \epsilon_1&=-\frac{\dot
H}{H^2},&\epsilon_2&=\frac{\ddot\phi}{H\dot\phi},&\epsilon_3&=\frac{\dot E}{2HE},&\epsilon_4&=\frac{\dot Q_{GB}}{2HQ_{GB}},\,
\end{align}
where the auxiliary parameters $E$ is defined as
$E=\frac{\alpha}{\kappa^2}\left(1+\frac{ 3Q_a^2}{2 Q_{GB}\dot \phi^2}\right)$ with $Q_a=-8\dot \xi H^2$. In this context, $E$ carries information about string corrections and as we shall showcase subsequently, it affects the scalar spectral index of primordial perturbations.

Our goal at this point is the construction of an inflationary cosmological model in the context of Einstein-Gauss-Bonnet gravity. The viability of the model depends on the latest observational constraints according to Ref. \cite{Planck:2018vyg}. Furthermore, we determine the numerical spectrum of the dimensionless parameter $\alpha$ in order to achieve compatibility with the Swampland criteria as well. The observational indices, namely the scalar spectral index of primordial perturbations $n_\mathcal{S}$, the tensor spectral index $n_\mathcal{T}$ and the tensor-to-scalar ratio r, are given by the following expressions,
\begin{align}
\label{observedGB}
\centering
n_\mathcal{S}&\simeq1-2(2\epsilon_1+\epsilon_2+\epsilon_3),&n_\mathcal{T}&\simeq-2(\epsilon_1+\epsilon_4),&r&\simeq16\left|\left(\alpha \epsilon_1-\frac{\kappa^2Q_e}{4H}\right)\frac{c_A^3}{\kappa^2Q_{GB}}\right|,\,
\end{align}
where the field propagation velocity $c_A$ is defined as,
\begin{equation}
\centering
\label{soundwaveGB}
c_\mathcal{A}^2=1+\frac{Q_aQ_e}{2Q_{GB}\dot \phi^2+3Q_a^2},\,
\end{equation}
with $Q_e=-32\dot\xi \dot H$ being an additional auxiliary parameter of the theory. According to the latest Planck data \cite{Planck:2018vyg}, the numerical value of the scalar spectral index of primordial perturbations is $n_\mathcal{S}=0.9649\pm 0.0042$ with 67$\%$ confidence level and the tensor-to-scalar ratio must be $r<0.064$ with 95$\%$ confidence level. Concerning the tensor spectral index, its value has not been determined yet, given that no B-modes have been observed so far however, this is a very exciting topic which may be resolved in the near future. It is worth mentioning that the inclusion of string corrective terms in the gravitational action has the advantage of affecting tensor perturbations in such a way that a blue tilted tensor spectral index may be extracted, see Ref. \cite{Venikoudis:2021oee,Fronimos:2021czc}. The models which shall be studied in the present article also share this feature however, the possibility of extracting a red tilted tensor spectral index as it is usual in the inflationary phenomenology of canonical scalar field is not excluded.

In the following, the methodology which is used for the evaluation of the observational indices during the first horizon crossing is presented. Firstly, we evaluate the final value of the scalar field at the end of the inflationary era by setting the first slow-roll index equal to unity. Then, the initial value of the inflaton can be extracted based on the equation of the $e$-folding number which is given by the expression, $N=\int_{t_i}^{t_f}{Hdt}=\int_{\phi_i}^{\phi_f}{\frac{H}{\dot\phi}d\phi}$, where the difference $t_f-t_i$ indicates the duration of the inflationary era. According to the definition of $\dot \phi$ in Eq. (\ref{fdot}), one can extract the $e$-folding equation as,
\begin{equation}
\label{efoldsGB}
N=\int_{\phi_i}^{\phi_f}\frac{\xi''}{\xi'}d\phi,
\end{equation}
which is assumed to be of near $N\sim50-60$. Let us now proceed with the numerical results for two models of interest. The models which shall be considered have been reviewed in Ref. \cite{Odintsov:2020sqy} for the case of $\alpha=1$ and are known for their viability.

\subsection{Model with Error function as the Gauss-Bonnet scalar coupling function}

We begin the numerical analysis by introducing the Gauss-Bonnet scalar coupling function, which is assumed to be
 \begin{equation}
 \centering
 \label{erf}
 \xi(\phi)=-Erf(\delta\kappa\phi)\, ,
 \end{equation}
 where $\delta$ is a dimensionless parameter. Such a choice is proven to be quite convenient as the ratio $\frac{\xi'}{\xi''}$, which appears in most of the expressions presented before is simplified to a great extend and reads $\frac{\xi'}{\xi''}=-\frac{1}{2 \delta ^2 \kappa ^2 \phi }$. It is essentially the inverse choice of the power-law model. Then, the scalar potential, according to the continuity equation (\ref{motion3GBfinal}) reads,
 \begin{equation}
 \centering
 \label{erfpot}
 V(\phi)=\frac{3 \sqrt{\pi } \alpha ^2 \phi ^{\frac{1}{2 \alpha  \delta ^2}} \left(\delta ^2 \kappa ^2 \phi ^2\right)^{\frac{1}{4} \left(\frac{1}{\alpha  \delta ^2}+2\right)}}{3 \sqrt{\pi } \alpha ^2 c \left(\delta^2 \kappa ^2 \phi ^2\right)^{\frac{1}{4} \left(\frac{1}{\alpha  \delta ^2}+2\right)}+8 \delta  \kappa ^5  \phi ^{\frac{1}{2 \alpha  \delta ^2}+1} \Gamma \left(\frac{1}{4} \left(2+\frac{1}{\alpha  \delta ^2}\right),\delta^2 \kappa ^2 \phi ^2\right)}\, ,
 \end{equation}
 where here, $c$ is the integration constant with mass dimensions $[c]=\ev^{-4+\frac{1}{2\alpha\delta^2}}$ for the sake of consistency. In turn, a direct replacement of the scalar potential and the Gauss-Bonnet scalar coupling function on the slow-roll indices  generates the following relations,
 
 \begin{equation}
 \centering
 \label{erfindex1}
 \epsilon_1=\frac{1}{8 \alpha  \delta ^4 \kappa ^2 \phi ^2}\, ,
 \end{equation}
 
 \begin{equation}
 \centering
 \label{erfindex2}
 \epsilon_2=\frac{4 \alpha  \delta ^2-1}{8 \alpha  \delta ^4 \kappa ^2 \phi ^2}\, .
 \end{equation}
Due to the intricacy of the scalar potential, only the first two slow-roll indices are showcased however, all of them are important for the inflationary phenomenology as we shall see later based on their numerical values. Afterwards, solving the equation $\epsilon_1(\phi_f)=1$ with respect to the final value of the scalar field $\phi_f$ suggests that such value depends on the free parameters of the model as shown below,

\begin{figure}[h!]
\centering
\includegraphics[width=15pc]{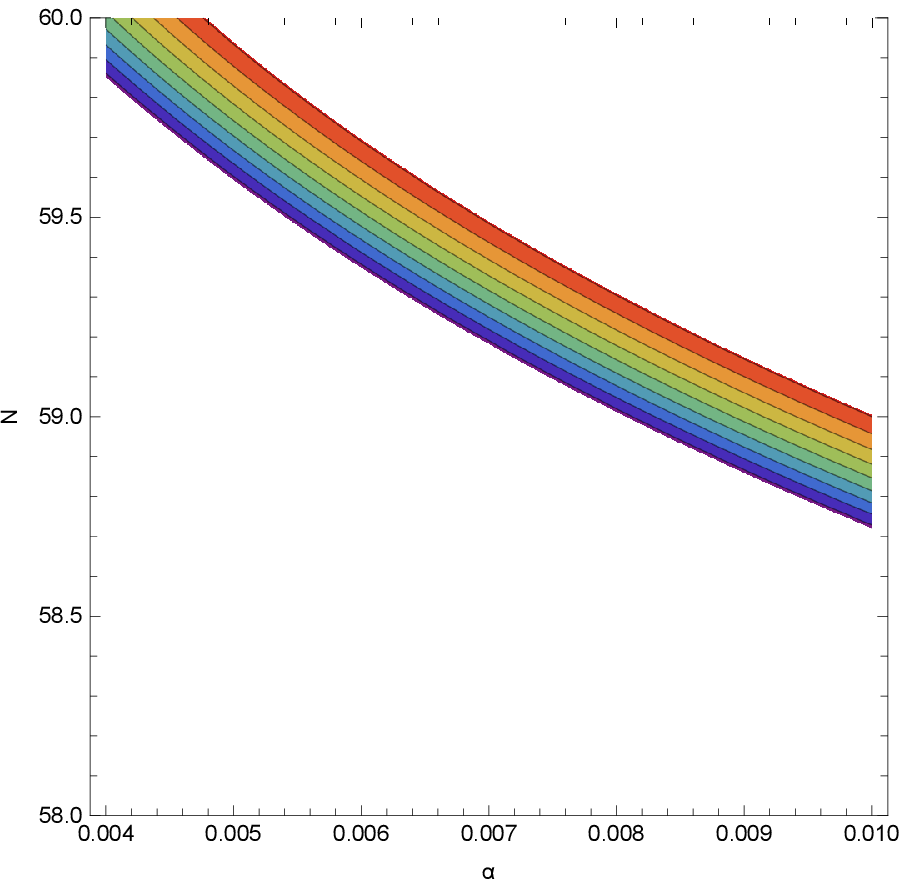}
 \includegraphics[width=3pc]{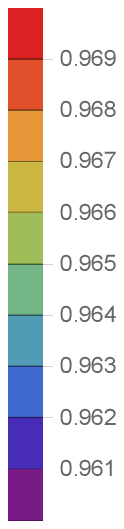}
 \caption{Scalar spectral index $n_\mathcal{S}$ of primordial perturbations for the error function model. As shown, the area of compatibility is quite narrow for pairs of $\alpha$ and $N$.}
 \label{nsmodel1}
 \end{figure}

 \begin{figure}[h!]
\centering
\includegraphics[width=15pc]{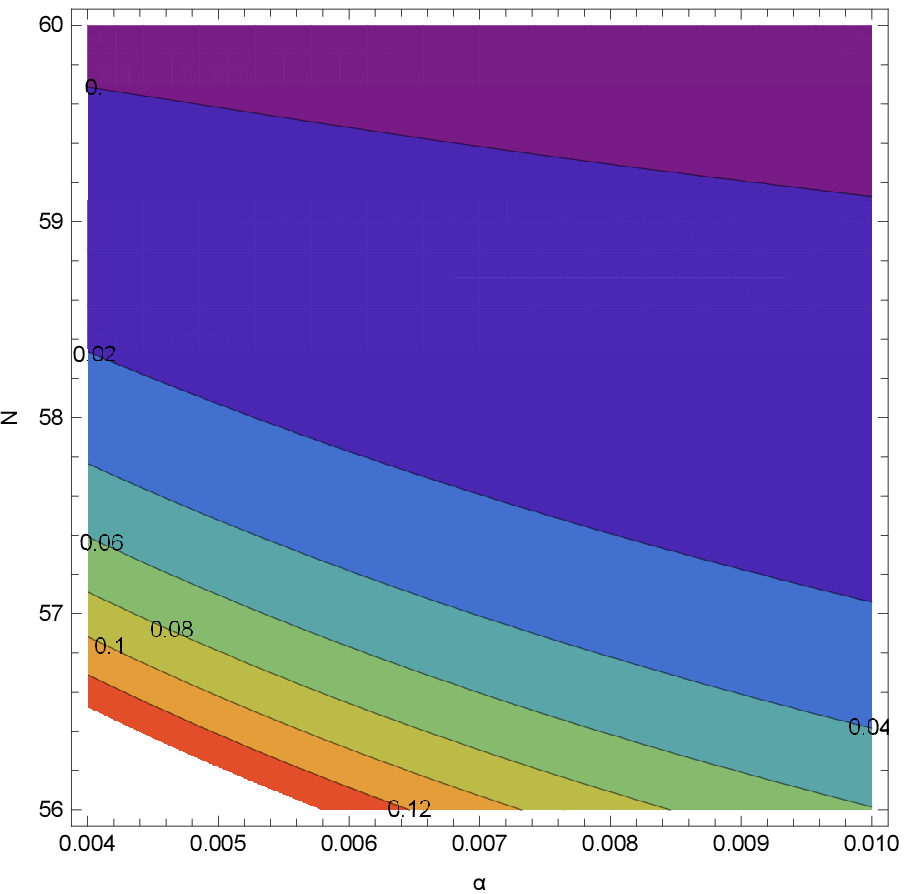}
 \includegraphics[width=15pc]{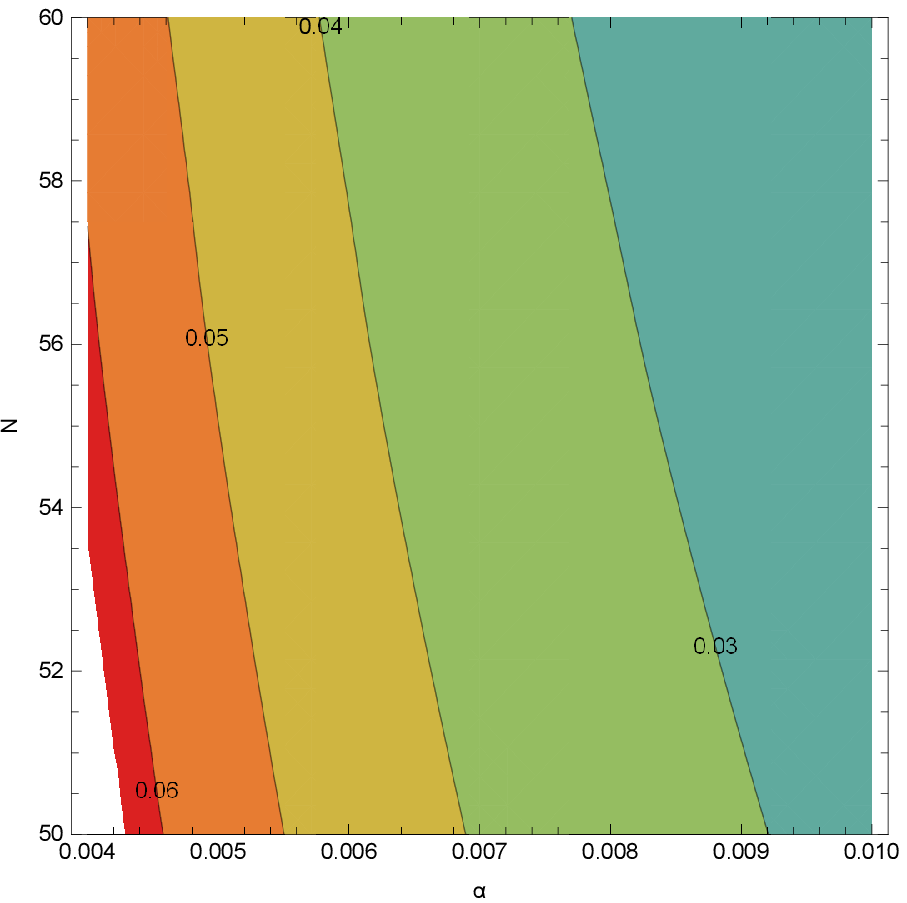}
 \caption{Contour plots of the tensor spectral index $n_\mathcal{T}$ (left) and tensor to scalar ratio $r$ (right) with respect to e-folding number and the rescale parameter $\alpha$. In contrast to the scalar spectral index, both indices have a wide range of viability.}
 \label{tensormodel1}
 \end{figure}
 
 \begin{equation}
 \centering
 \label{phiferf}
 \phi_f=\frac{1}{2 \sqrt{2} \sqrt{\alpha } \delta ^2 \kappa }\, ,
 \end{equation}
 where it becomes abundantly clear that such value is affected greatly by the choice of the Gauss-Bonnet scalar coupling function. In turn, not only the observables $n_\mathcal{S}$, $r$ and $n_\mathcal{T}$ but also the Swampland criteria which are calculated during the first horizon crossing are affected strongly by such scalar function and not only through the scalar potential. This is an additional dependence since now both $\phi_i$ and $V(\phi)$ are connected to $\xi(\phi)$. Now, by solving the Eq. (\ref{efoldsGB}) with respect to $\phi_i$ one finds that
 
 \begin{equation}
 \centering
 \label{phiierf}
 \phi_i=\frac{\sqrt{\frac{1}{8 \alpha  \delta ^2}+N}}{\delta  \kappa }\, ,
 \end{equation}
 where, as expected, it depends on $\alpha$, $\delta$ and the e-foldings number $N$. Hence, we now have everything we need in order to ascertain the validity of the model and examine the Swampland criteria. Note that in order for the model to be a viable choice, the sound wave velocity should also be calculated during the first horizon crossing in order to examine whether the model suffers from ghost instabilities. All examples which are considered in this section, as shown below, are indeed free of instabilities.
 
 Let us now proceed with the extraction of the numerical value of the observed indices. For the sake of simplicity we shall assume that the Planck mass is equal to unity therefore $\kappa=1$ hereafter. By designating $\left(N, c, \alpha, \delta\right)=\left(60, 1.1\cdot10^{-20}, 0.004, 12\right)$, the choice of the error function as a Gauss-Bonnet scalar coupling function seems to generate results that are compatible with observations as the scalar spectral index obtains the value $n_\mathcal{S}=0.964659$, the tensor to scalar ratio is now equal to $r=0.0576594$ and finally the tensor spectral index is expected to be $n_\mathcal{T}=-0.00191182$. It is worth mentioning that the inclusion of string corrections does not alter the sign of the tensor spectral index however now $r\simeq-30n_\mathcal{T}$. This result is purely numerical and not common among constraint Gauss-Bonnet models. The integration constant $c$ needs to obtain a value of order $\mathcal{O}\left(10^{-20}\right)$ in order to obtain a viable scalar spectral index. The result seems to differ from the one obtained in \cite{Odintsov:2020sqy} due to the inclusion of the rescale parameter $\alpha$. For the sake of generality, the scalar spectral index is depicted in Fig.\ref{nsmodel1} while, the tensor spectral index and the tensor to scalar ratio are available in Fig.\ref{tensormodel1}. It is important to highlight that the scalar spectral index has a rather small range of compatibility compared to the rest of the observables. Note also that the sound wave velocity reads $c_\mathcal{A}=0.999986$, thus justifying the previous statement on the absence of instabilities.

\begin{figure}[h!]
\centering
\includegraphics[width=15pc]{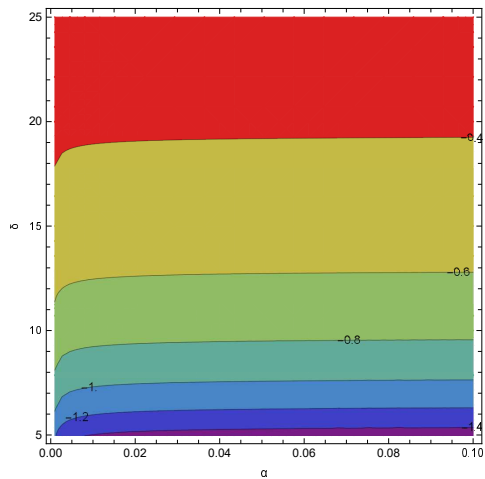}
 \caption{$\kappa\Delta\phi$ difference depending on $\alpha$ and $\delta$ for the error function model. The scalar coupling function seems to have a major impact on the fist condition. This is because the final value of the scalar field $\phi_f$ is specified from the first slow-roll index which in this context depends on the ratio $\xi'/\xi''$ and $\phi_i$ is specified from $\phi_f$.}
 \label{swamp1model1}
 \end{figure}
\begin{figure}[h!]
\centering
\includegraphics[width=15pc]{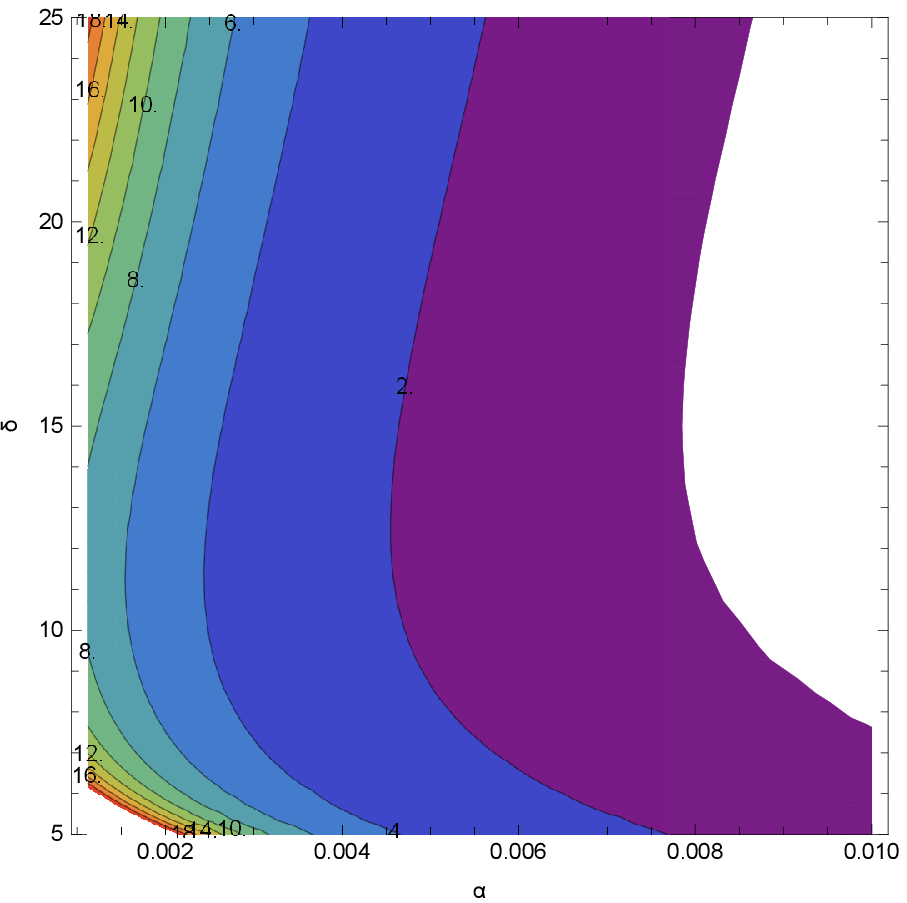}
 \includegraphics[width=15pc]{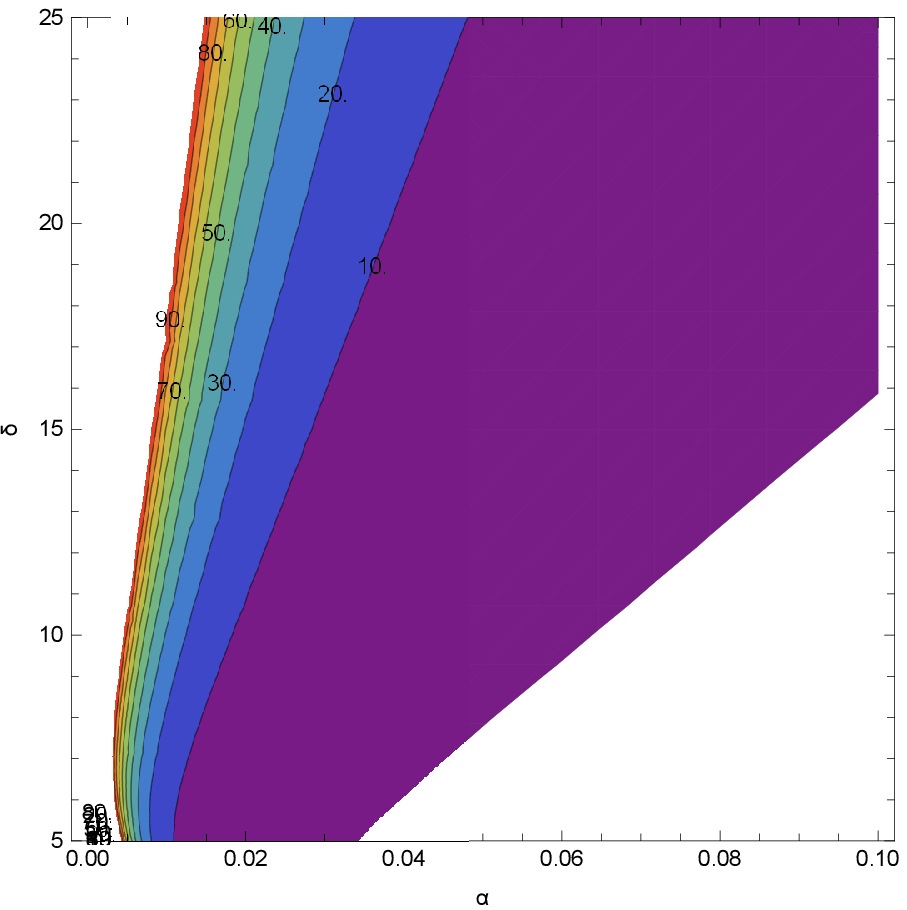}
 \caption{Second and third criteria $\frac{V'(\phi_i)}{\kappa V(\phi_i)}$ and $-\frac{V''(\phi_i)}{\kappa^2V(\phi_i)}$ on the left and right diagram respectively. As depicted, both parameters $\alpha$ and $\delta$ influence such ratios and there exists a wide range of pairs of values that satisfies both conjectures.}
 \label{swamp2model1}
 \end{figure}

 For the Swampland criteria, one can easily observe that the aforementioned set of parameters suggests that $\kappa\phi_i=0.646664$ and $\kappa\phi_f=0.0388206$. In this case the scalar field decreases with time and as shown, $\kappa\Delta\phi=-0.607843$ therefore the first condition is satisfied. In Fig.\ref{swamp1model1}, the dependence of $\kappa\Delta\phi$ on parameters $\delta$ and $\alpha$ is shown. It becomes clear that the free parameter of the Gauss-Bonnet scalar coupling function affects the first criterion strongly compared to the rescale parameter $\alpha$. Hence, the model at hand respects the Swampland criteria. Concerning the rest of the conditions, we mention that during the first horizon crossing $\frac{V(\phi_i)}{\kappa V(\phi_i)}=2.32758$ and $-\frac{V''(\phi_i)}{\kappa^2V(\phi_i} =177.851$ which are both greater than $\mathcal{O}(1)$ thus we conclude that all conditions are met. A more detailed analysis is shown in Fig.\ref{swamp2model1}. It is worth noting that the choice of $\xi(\phi)=\erf(\delta\kappa\phi)$ for the exact same set of parameters manages to affect the conjectures to a great extent. This is because of the constraint in the propagation velocity of tensor perturbations that was imposed in the model at hand, as $\dot\phi=H\frac{\xi'}{\xi''}$ under the slow-roll assumption, that influences the expression for the scalar potential through the continuity equation of the scalar field. In particular, the second and third conditions obtain a different sign and the ratio $\frac{V'(\phi_i)}{\kappa V(\phi_i)}$ becomes of order $\mathcal{O}\left(10^{-1}\right)$. This is not limited to just a sign change but in fact the form of the scalar coupling function.
 
 As a final step, we ascertain the validity of the slow-roll conditions for the error function model. For the set of free parameters selected previously, the numerical values of the slow-roll indices read $\epsilon_1=0.00360386$, $\epsilon_2=0.00469944$, $\epsilon_3=0.0057635$ and $\epsilon_4=-0.00264795$. Therefore, the slow-roll conditions are satisfied despite the fact that $c$ obtains such a small value. This results in the increase of the order of magnitude of the scalar potential which now becomes approximately $V(\phi_i)\simeq6.2\cdot10^{19}$, in reduced Planck units. Choosing different values, for instance $c=1$ suggests that $V\sim\mathcal{O}(1)$ but the scalar spectral index is not within the range $0.9649\pm0.0042$. Also, altering the rest of the parameters seems to affect severely the tensor to scalar ratio.

\subsection{Model with exponential-like scalar coupling function $\xi(\phi)$}

Suppose now that the Gauss-Bonnet scalar coupling function is defined as,
\begin{equation}
\centering
\label{couplingGB}
\xi(\phi)=\lambda \int^{\kappa\phi} e^{-\gamma x^m}dx,
\end{equation}
where $x$ is an auxiliary integration variable, while $\lambda$ and $\gamma$ are dimensionless free parameters. This function can be interpreted as a string correction thus, numerically speaking, has negligible contribution on the overall phenomenology. However, the ratio $\frac{\xi'}{\xi''}$ has a very important role due to its involvement in the e-folding equation. Specifically, notice that the ratio,
\begin{equation}
\centering
\label{ratio}
\frac{\xi'}{\xi''}=-\frac{\phi  (\kappa  \phi )^{-m}}{\gamma  m},
\end{equation}
has a power-law form. Based on the equation of the continuity of the scalar field (\ref{motion3GBfinal}), one can treat such an equation as a first order differential equation with respect to the scalar potential and extract the solution which has the form,
\begin{equation}
\centering
\label{scalarpotentialGB}
V(\phi)=V_1 e^{\frac{(\kappa  \phi )^{2-m}}{\alpha  \gamma m(2- m)}},
\end{equation}
where $[V_1]=\ev^4$. Here, the factor $24\xi'H^4$ was neglected because, as we shall see later, $\gamma$ has to be quite large in order to obtain a viable inflationary era and, consequentially, the value of $\xi$ and their derivatives are quite small. Nevertheless, the ratio $\frac{\xi'}{\xi''}$ is still important. In this numerical approach, the exponent $m$ is not strictly an integer therefore, the potential can have any dependence on the scalar field. Note also that the Gaussian case corresponding to $m=0$ for the scalar potential and $m=2$ for the Gauss-Bonnet scalar coupling function are essentially excluded. The case of $m=0$ in this approach would correspond to a linear Gauss-bonnet scalar coupling function which has been proven to be incompatible with the Planck data, since the constraint on the propagation velocity of tensor perturbations predicts a constant-roll condition $\ddot\phi=H\dot\phi$ for the scalar field, an interested reader may refer to \cite{Oikonomou:2020oil}.

Let us now proceed with the deduction of the values of the scalar field during the initial and final stage of the inflationary era. To do so, one needs to extract information about the auxiliary parameters. In particular, the first two slow-roll indices of the model at hand are given by the following equations,

\begin{figure}[h!]
\centering
\includegraphics[width=15pc]{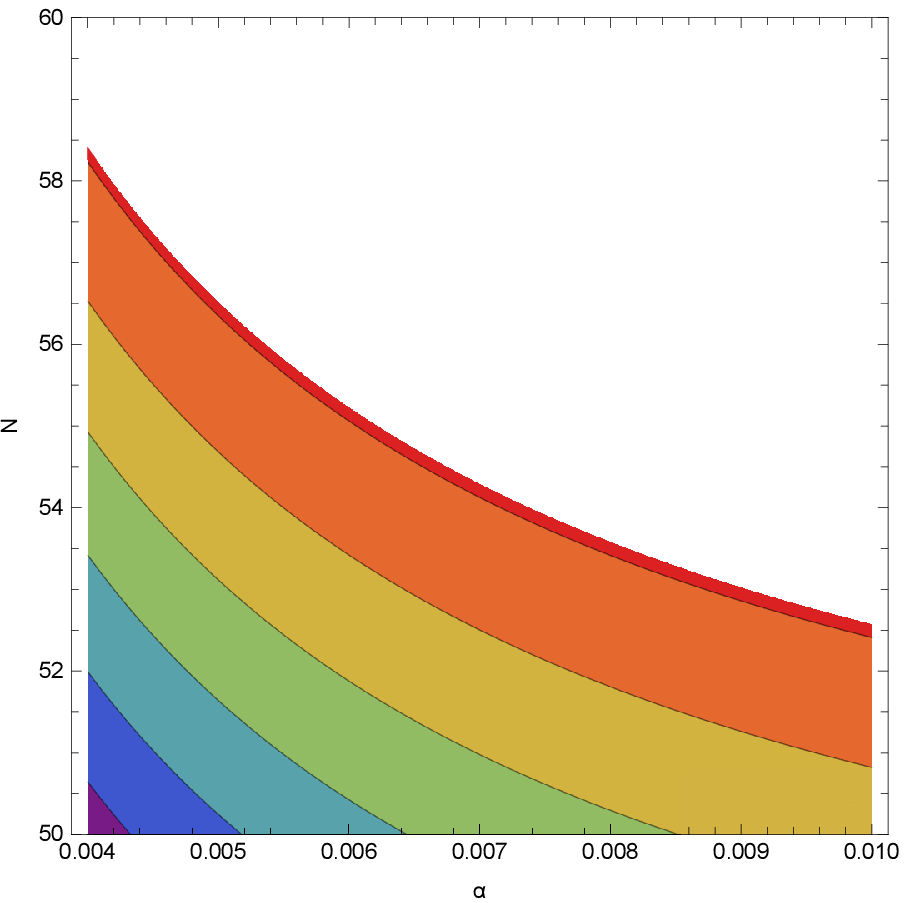}
\includegraphics[width=4pc]{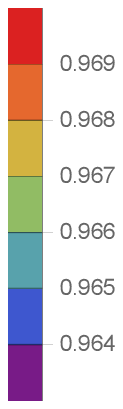}
\caption{Scalar spectral index $n_\mathcal{S}$ as a function of $\alpha$ and $N$. Here, compatibility is achieved for small values of $\alpha$ assuming that the e-folding number resides in the area [50, 60]. }
\label{nsGB}
\end{figure}

\begin{equation}
\centering
\label{epsilon1GB}
\epsilon_1\simeq \frac{(\kappa  \phi )^{2-2 m}}{2 \alpha  \gamma ^2 m^2},
\end{equation}

\begin{equation}
\centering
\label{epsilon2GB}
\epsilon_2\simeq \frac{(\kappa  \phi )^{-2 m} \left(2 \alpha  \gamma  (m-1) m (\kappa  \phi )^m-\kappa ^2 \phi ^2\right)}{2 \alpha  \gamma ^2 m^2},
\end{equation}
where indices $\epsilon_3$ and $\epsilon_4$ have been omitted because of their perplexed forms. However, their numerical value is expected to be quite small compared to the rest. In consequence,  by setting the first slow-roll index 
equal to unity, the final value of the inflaton reads 
\begin{equation}
\centering
\label{phifinalGB}
\phi_f=\frac{\left(2 \alpha  \gamma ^2 m^2\right)^{\frac{1}{2-2 m}}}{\kappa }\, .
\end{equation}

\begin{figure}[h!]
\centering
\includegraphics[width=15pc]{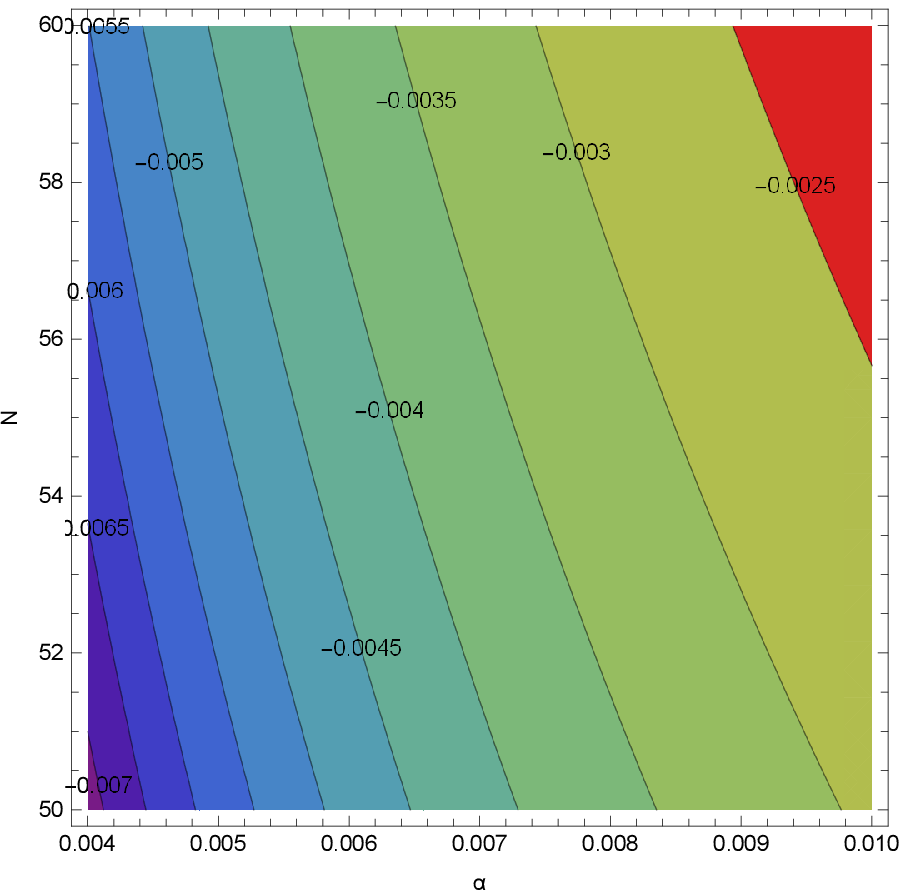}
\includegraphics[width=15pc]{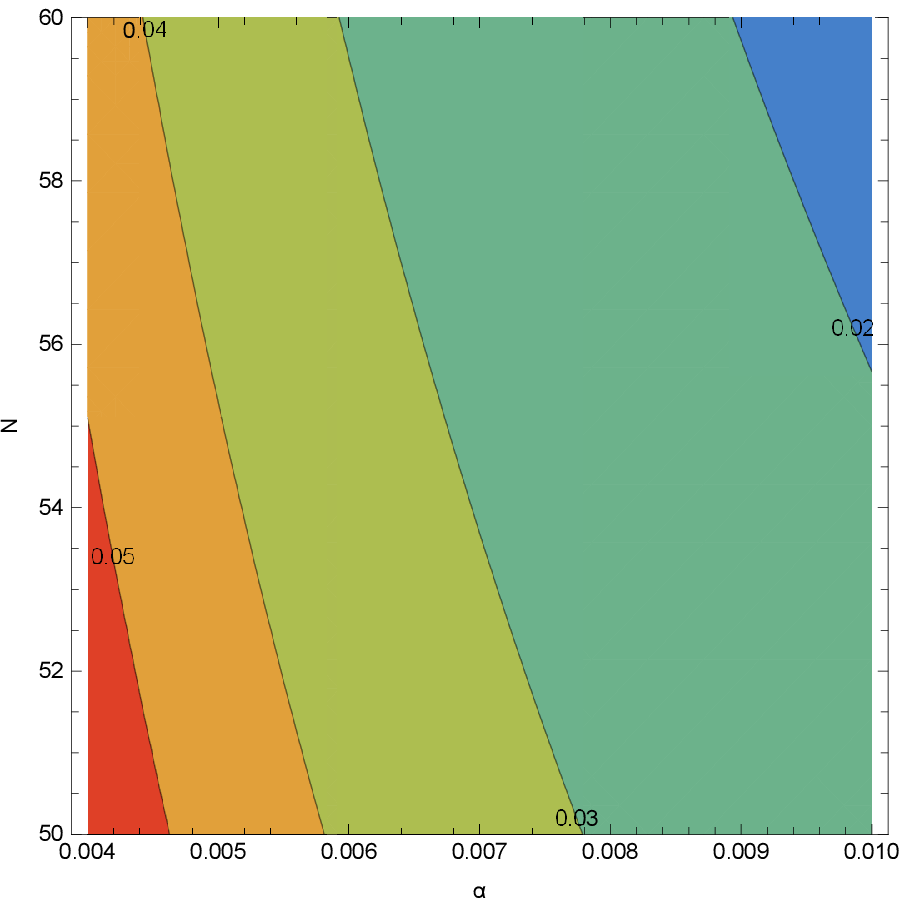}
\caption{Tensor spectral index (left) and tensor to scalar ratio dependence on the rescale parameter $\alpha$ and the e-folding number $N$. It becomes abundantly clear that while the model is indeed compatible with observations for a plethora of values, the tensor spectral index is expected to be red tilted for small values of parameter $\alpha$.}
\label{tensorGB}
\end{figure}

\begin{figure}[h!]
\centering
\includegraphics[width=15pc]{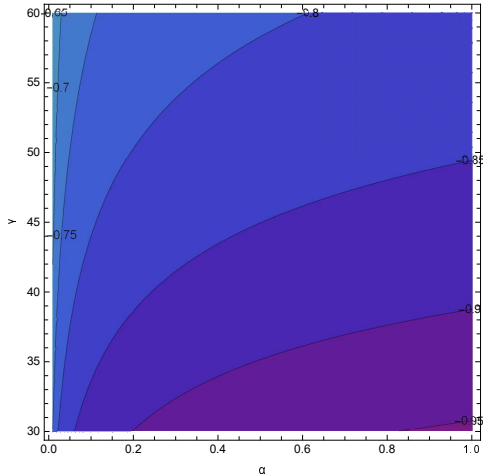}
\caption{$\kappa\Delta\phi$ difference for the Gauss-Bonnet model. In this case, one can easily see that the criterion is satisfied for large values of the parameter $\gamma$ which is in agreement with the conditions that render the scalar spectral index compatible with observations.}
\label{swamp1GB}
\end{figure}

In principle there exist several other values of the scalar field that satisfy the condition $\epsilon_1(\phi_f)=1$. However, we shall limit our work to only the aforementioned value which is real for the case of a non negative $\alpha$. The substitution of the aforementioned value into the upper limit of the e-folding equation leads to the initial value of the scalar field during the first horizon crossing
\begin{equation}
\centering
\label{phiinitialGB}
\phi_i=\frac{\left[\left(2^{\frac{1}{2-2 m}} \left(\alpha  \gamma ^2 m^2\right)^{\frac{1}{2-2 m}}\right)^m+\frac{N}{\gamma }\right]^{1/m}}{\kappa }.
\end{equation}
At this stage, let us specify the free parameters of the model in reduced Planck units, where $\kappa^2=1$ in order to ascertain the compatibility with the latest Planck data \cite{Planck:2018vyg}. Assuming that (N, $V_1$, $\lambda$, $\gamma$, m, $\alpha$)=$\left(50, M_P^4, -1, 35, 4, 0.004\right)$ the model produces results consistent with the latest Planck data. Specifically, the scalar spectral index assumes the value $n_\mathcal{S}=0.963497$ and the tensor-to-scalar ratio is $r=0.0576666$. Last but not least, the tensor spectral index of primordial perturbations takes the value $n_\mathcal{T}=-0.00720833$, indicating of slightly red-shifted spectrum. It becomes apparent that the relation $r=-8n_\mathcal{T}$ holds even though string corrections are present, since the Gauss-Bonnet scalar coupling function has a small contribution, however it affects $\dot\phi$. The scalar spectral index is essentially viable for a variety of values as shown in Fig.\ref{nsGB} where the e-folding number $N$ and the rescale parameter $\alpha$ were considered to be free while the rest was fixed, as suggested above. Here, as shown, both parameters influence the scalar spectral index to a great extent but only a small value of $\alpha$ should be considered in order to obtain a viable description. Similarly, in Fig.\ref{tensorGB} the tensor to scalar ratio and the tensor spectral index are depicted as functions of the same parameters. As mentioned above, the tensor spectral index is red tilted for a wide range of values. 
\begin{figure}[h!]
\centering
\includegraphics[width=15pc]{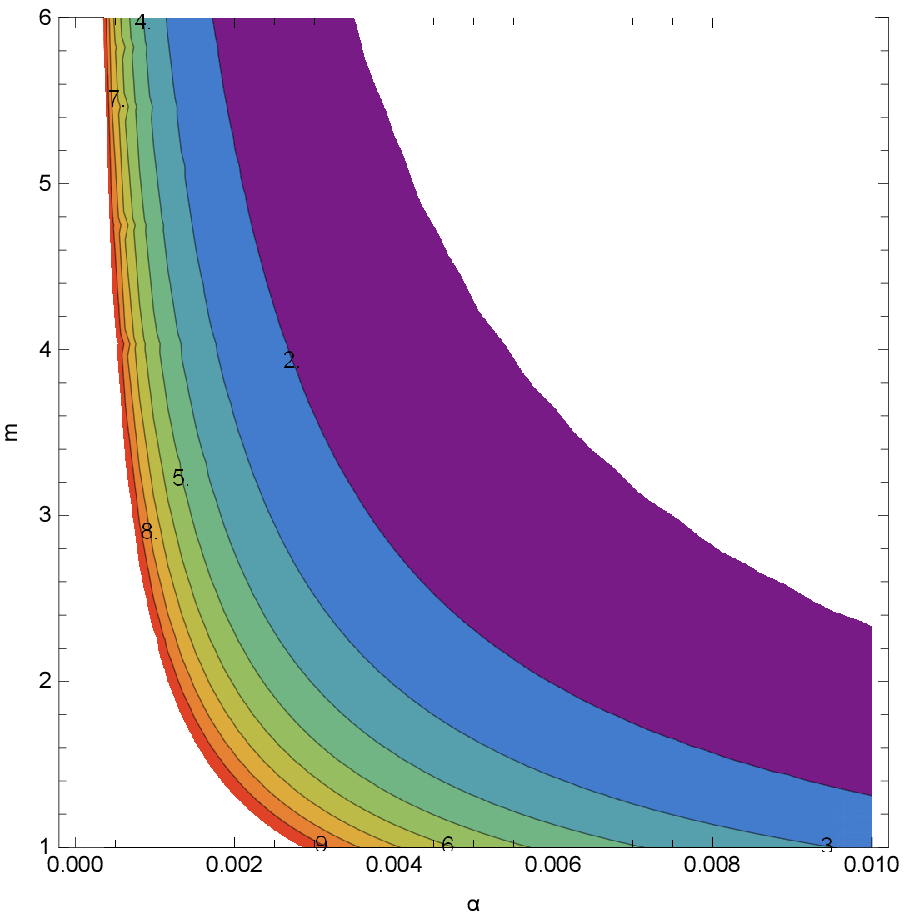}
\includegraphics[width=15pc]{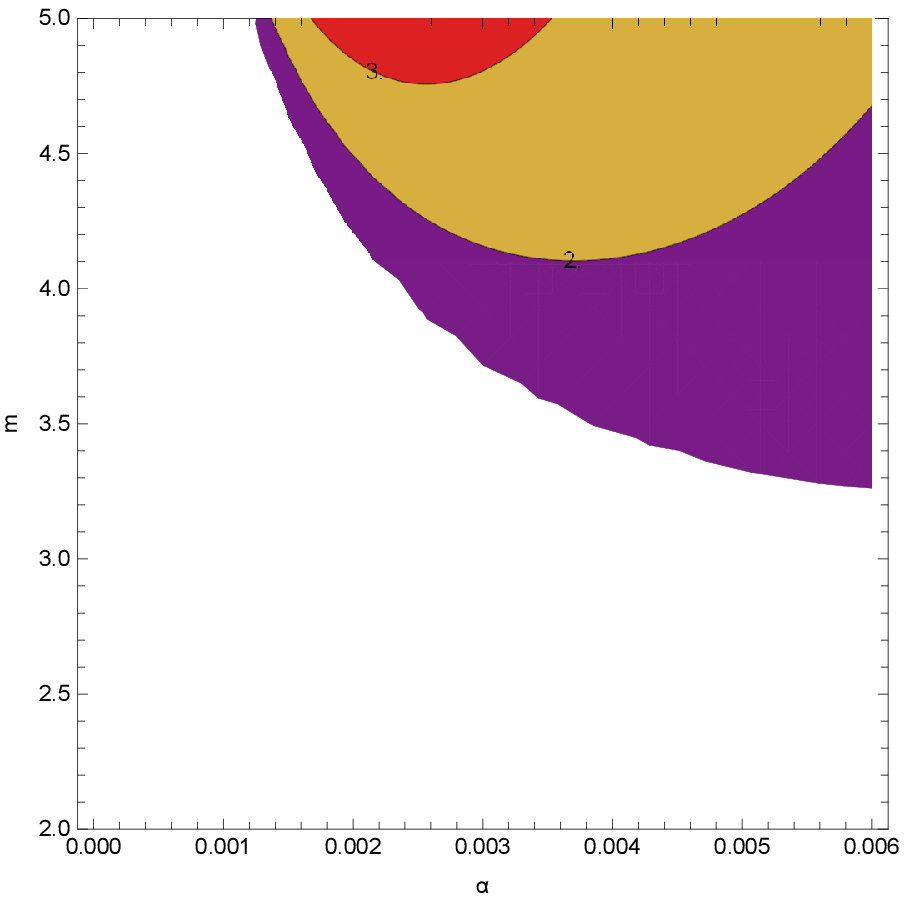}
\caption{Dependence of the$\frac{V'(\phi_i)}{\kappa V(\phi_i)}$ and $-\frac{V''(\phi_i)}{\kappa^2V(\phi_i)}$ ratios on parameters $m$ and $\alpha$ for the second Gauss-Bonnet model. As shown, they can be satisfied simultaneously.}
\label{swamp2GB}
\end{figure}
Concerning the inflaton, the initial and the final values during the inflationary era are $\kappa\phi_i=1.09979$ and $\kappa\phi_f=0.430635$ respectively, where  $|\kappa\Delta\phi|<\mathcal{O}(1)$ as suggested by the first criterion. This indicates that during the inflationary era the value of the homogeneous scalar field decreases. By examining the difference $\kappa\Delta\phi$ in Fig.\ref{swamp1GB}, one can easily see that the Swampland criteria are satisfied for small values of $\alpha$. The numerical analysis indicates that out of the three conditions examined in the present study, all of them are simultaneously met as shown in Fig.\ref{swamp1GB}. In particular, for the parameters indicated above, one can easily ascertain that $\frac{V'(\phi_i)}{\kappa V(\phi_i)}= 1.34242$ and $-\frac{V''(\phi_i)}{\kappa^2V(\phi_i)}= 1.85977$. Hence, all conditions are satisfied at the same time for the same set of parameters thus the Swampland criteria, for the model at hand, are satisfied as well. This feature, which is shared in both models, seems to validate the results found in Ref \cite{Odintsov:2020zkl} for the constrained Gauss-Bonnet model. Therefore, one can easily see that for the tensor to scalar ratio we obtain $r\simeq16\epsilon_1$ in both models. Additionally, during the initial stage of inflation, the slow-roll indices take the following numerical values, $\epsilon_1 \simeq 0.00129421$, $\epsilon_2 \simeq 0.0135276$, $\epsilon_3 \sim \mathcal{O}\left(10^{-39}\right)$ and $\epsilon_4\sim \mathcal{O}\left(10^{-20}\right)$. It is also worth noting that the model is free of ghost instabilities and respects causality since the sound wave velocity is equal to unity.

Furthermore, in order to validate the viability of the model, we shall investigate whether the slow-roll approximations (\ref{approx}) of the model are valid during the first horizon crossing, something which is hinted above from the numerical values of the slow-roll indices. By substituting $\phi$ with $\phi_i$, we get that 
$\dot H \sim \mathcal{O}\left(10^{-1}\right)$ and $H^2 \sim \mathcal{O}(10)$. Thus, the slow-roll condition holds true. Moreover,
the kinetic term of the scalar field is numerically negligible compared with the scalar potential, $\frac{1}{2}\dot \phi^2 \sim \mathcal{O}\left(10^{-4}\right)$, $V \sim \mathcal{O}\left(10^{-1}\right)$. Hence, the condition $\frac{1}{2}\dot \phi^2\ll V(\phi)$ is satisfied. Lastly, $\ddot \phi \sim \mathcal{O}\left(10^{-3}\right)$ and $H \dot \phi \sim \mathcal{O}\left(10^{-1}\right)$, concluding that the set of approximations in (\ref{approx}) are accurate. Concerning the Gauss-Bonnet string corrections, the term $24\xi' H^4 \sim \mathcal{O}\left(10^{-18}\right)$ has non-considerable numerical contribution to the equations of motion compared to $V'\sim \mathcal{O}\left(10^{-1}\right)$, concluding that the continuity equation of the inflaton is satisfied. In addition, $16\dot \xi H \dot H\sim  \mathcal{O}\left(10^{-23}\right)\ll\frac{1}{2}\dot \phi^2$ and $24\dot\xi H^3\sim\mathcal{O}\left(10^{-20}\right)$ is also insignificant compared to $V\sim \mathcal{O}(10^{-1})$.

\section{Theoretical Framework of Rescaled $f(R)$ Einstein-Chern-Simons gravity}
Let us now study a different rescaled model. In this approach we shall study the impact of string axionic term that affects tensor perturbations but leaves the equations of motion the same as the canonical scalar field. For the case at hand, the Chern-Simons model reads,
\begin{equation}
\centering
\label{actionCS}
\mathcal{S}=\int{d^4x\sqrt{-g}\left(\frac{\alpha R}{2\kappa^2}-\frac{1}{2}g^{\mu\nu}\nabla_\mu\phi\nabla_\nu\phi-V(\phi)+\frac{1}{8}\nu(\phi)R \Tilde{R}\right)}\, ,
\end{equation}
where now the new curvature invariant has the form $R\Tilde{R}=\epsilon^{\mu \nu \sigma \rho}R^{\ \ \alpha \beta }_{\mu \nu}R_{\sigma \rho \alpha \beta}$. The term $\epsilon^{\mu \nu \sigma \rho}$ expresses the totally antisymmetric Levi-Civita tensor in four dimensions and $\nu({\phi})$ indicates the Chern-Simons scalar coupling function. The implementation of the variational principle into the gravitational action (\ref{actionCS}) with respect to the metric tensor generates the field equations,
\begin{equation}
\label{fieldequations}
\centering
\alpha\bigg(R_{\alpha \beta}-\frac{1}{2}Rg_{\alpha \beta}\bigg)=\kappa^2\bigg[\nabla_{\alpha}\phi\nabla_{\beta}\phi-\frac{1}{2}g^{\mu \nu}\nabla_{\mu}\phi\nabla_{\nu}\phi g_{\alpha \beta}+\epsilon_{\alpha}^{\ \mu \nu \sigma}\bigg(\nabla_{\rho}\nabla_{\sigma}\nu(\phi)R^{\rho}_{\ \beta\mu \nu}-2\nabla_{\sigma}\nu (\phi) \nabla_{\nu}R_{\beta \mu}\bigg)\bigg]
\end{equation}
and by isolating the time component, the spatial component and the continuity equation of the scalar field can easily ascertain that the equations of motion are indeed exactly the same as in the case of the canonical scalar field, that is

\begin{equation}
\centering
\label{motion1CS}
\frac{3\alpha H^2}{\kappa^2}=\frac{1}{2}\dot\phi^2+V,\,
\end{equation}
\begin{equation}
\centering
\label{motion2CS}
-\frac{2 \alpha \dot H}{\kappa^2}=\dot\phi^2,\,
\end{equation}
\begin{equation}
\centering
\label{motion3CS}
\ddot\phi+3H\dot\phi+V'=0.\
\end{equation}

Before we proceed with the slow-roll phenomenology for the model, it is interesting to investigate the impact of the Chern-Simons term in the theory regardless of its absence in the equations of motion. As it is already mentioned, the addition of the term $\sim \nu(\phi)R\tilde{R}$ in the gravitational action (\ref{actionCS}) affects the tensor perturbations and in consequence the tensor spectral index $n_\mathcal{T}$ and the tensor to scalar ratio $r$. Based on \cite{Hwang:2005hb}, the tensor perturbations are described by the following expression,
\begin{equation}
\label{tensorperturbations}
\centering
\frac{1}{a^3}\frac{d}{dt}\bigg(a^3 \dot C_{\alpha \beta}\bigg)-\frac{\nabla^2}{a^2}C_{\alpha \beta}-\frac{2k^2}{a}\epsilon_{(\alpha}^{\ \ \mu \nu}\bigg[(\ddot\nu-H\dot\nu)\dot C_{\beta)\mu}+\dot\nu D_{\beta)\mu}\bigg]_{,\nu}=0,
\end{equation}
where $D_{\alpha\beta}=\Ddot{C}_{\alpha \beta}+3H\dot{C}_{\alpha \beta}-\frac{\nabla^2}{\alpha^2}C_{\alpha \beta}$.
By making the ansatz
\begin{equation}
\centering
\label{ansatz}
C_{\alpha \beta}=\sqrt{Vol}\int \frac{d^3k}{(2\pi)^3}\sum_{l}\epsilon_{\alpha\beta}^{(l)}(\Vec{k})h_{l\Vec{k}}e^{i\Vec{k}\cdot\Vec{x}},
\end{equation}
such that we can split the tensor modes into left-handed and right-handed polarizations (where the l-index is summed over them), the Eq. (\ref{tensorperturbations}) is diagonalized as follows,
\begin{equation}
    \centering
    \label{modeeq}
    \frac{1}{a^3Q_{CS}}\frac{d}{dt}\left(a^3Q_{CS}\dot h_{l \vec{k}}\right)+c_\mathcal{T}^2\frac{k^2}{a^2}h_{l \vec{k}}\, ,
\end{equation}
where $Q_{CS}=\frac{\alpha}{\kappa^2}+2\lambda_l\dot \nu \frac{k}{a}$. We highlight that the parameter $h_{l\vec{k}}$ corresponds to the tensor perturbation of a certain polarization l for a given mode k, whereas $\epsilon_{\alpha \beta}^{(l)}$ is the polarization tensor. Furthermore, $\lambda_l$ is an auxiliary parameter which takes the values $\lambda_L=-1$ and $\lambda_R=1$ for left-handed and right-handed polarizations, respectively. Hence, according to the above reasoning, the auxiliary parameter $Q_{CS}$ affects the tensor perturbations. Consequently, the tensor spectral index and the tensor to scalar ratio are affected by the Chern-Simons scalar coupling function $\nu(\phi)$, whereas the spectral index of scalar perturbations remains unaffected. We also note that the propagation velocity of tensor perturbations in this context is not affected, meaning that $c_\mathcal{T}=1$ thus the coefficient of $h_{l\vec{k}}$ in Eq.(\ref{modeeq}) is the ratio $\frac{k^2}{a^2}$. Let us now proceed with the slow-roll phenomenology by investigating the impact of the Chern-Simons term in the slow-roll indices and in the observational indices. Considering that the slow-roll conditions (\ref{approx}) during the inflationary era hold true, the resulting equations of motion are,
\begin{equation}
\centering
\label{motion1CSfinal}
H^2\simeq \frac{\kappa^2 V}{3\alpha},\,
\end{equation}
\begin{equation}
\centering
\label{motion2CSfinal}
\dot H=-\frac{\kappa^2\dot \phi^2}{2\alpha},\,
\end{equation}
\begin{equation}
\centering
\label{motion3CSfinal}
\dot \phi \simeq -\frac{V'}{3H}.\
\end{equation}
The cosmological dynamics of the primordial era is described by the slow-roll parameters; namely
\begin{align}
\centering \epsilon_1&=-\frac{\dot
H}{H^2},&\epsilon_2&=\frac{\ddot\phi}{H\dot\phi},&\epsilon_3&=\frac{1}{2}\sum_{L,R}\frac{\dot Q_{CS}}{2HQ_{CS}}.\,
\end{align}
Based on the slow-roll parameters it is obvious that the Chern-Simons gravitational term affects only the index $\epsilon_3$. Specifically due to the parity violation, left handed and right handed polarization states of gravitational waves have a different impact on tensor perturbations.

Let us now see how the observed indices, namely the scalar spectral index of primordial curvature perturbations, the tensor spectral index and the tensor to scalar ratio, differ from the Gauss-Bonnet case. Due to the fact that the Chern-Simons term, as expressed previously, affects only tensor perturbations, the scalar spectral index is the same as in the canonical scalar field case while the auxiliary parameter $Q_{CS}$ and its derivative have an influence on the tensor spectral index and the tensor to scalar ratio as shown below 

\begin{align}
\label{observedCS}
\centering
n_\mathcal{S}&=1-2(2\epsilon_1+\epsilon_2),&n_\mathcal{T}&=-2(\epsilon_1+\epsilon_3),&r&=8\alpha\epsilon_1\sum_{L,R}\frac{1}{\kappa^2 |Q_{CS}|}.\,
\end{align}
Note that the free parameter $\alpha$ is strictly positive in this approach therefore, the tensor to scalar ratio is safely defined. Note also that when the Chern-Simons scalar coupling function is subleading compared to $M_P^2$, the numerical value of $n_{\mathcal{T}}$ and $r$ is quite close to the one obtained in the case of the canonical scalar field. It is worth mentioning  that whenever the condition $\epsilon_3<-\epsilon_1$ is satisfied, the model predicts a blue tilted tensor spectral index. Indeed, as we shall showcase in the subsequent models, one can achieve such a feature for several cases. This is an interesting outcome since a blue tilted tensor spectral index can result in the enhancement of the amplitude of primordial gravitational waves, see Ref \cite{Odintsov:2021kup} for further details. Such models may be in position to explain a possible gravitational wave detection in the near future.

For the following models, the previous approach shall be considered once again. In particular, the initial value of the scalar field during the first horizon crossing will be extracted from the e-foldings number

\begin{equation}
\centering
\label{efolds2}
N=\int_{\phi_f}^{\phi_i}{\frac{\kappa^2 V}{\alpha V'}\mathrm{d}\phi},
\end{equation}
which differs from the canonical case by a factor of $\frac{1}{\alpha}$, as expected from the rescaled model. The initial value of the scalar field shall be used as an input in both the observed indices and the scalar potential in order to achieve compatibility with the Planck 2018 collaboration data and, in addition, ascertain the validity of the Swampland criteria.

\subsection{Model with Power-law scalar coupling function}
The first model we focus on involves a scalar field that rolls slowly in a scalar potential of the form
\begin{equation}
\centering
\label{scalarpotentialpowerlaw}
V(\phi)=V_2(\kappa \phi)^n,
\end{equation}
where $[V_2]=\ev^4$ for consistency reasons. The scalar coupling function has the following simple form
\begin{equation}
\label{couplingpowerlaw}
\centering
\nu(\phi)=(\kappa \phi )^p,
\end{equation}
where $p$ is, for the time being, arbitrary and in principle should not be considered as an integer. The first two gravitational equations of motion for the model can be written as
\begin{equation}
\centering
\label{equation1powerlawCS}
H^2 \simeq \frac{V_2\kappa^2 (\kappa \phi)^n}{3\alpha},
\end{equation}
\begin{equation}
\centering
\label{equation2powerlawCS}
\dot H \simeq - \frac{n^2 V_2 (\kappa \phi)^n}{6\phi^2}.
\end{equation}
The slow-roll parameters determine the cosmological evolution during the primordial era and are given by the following expressions

\begin{figure}[h!]
\centering
\includegraphics[width=15pc]{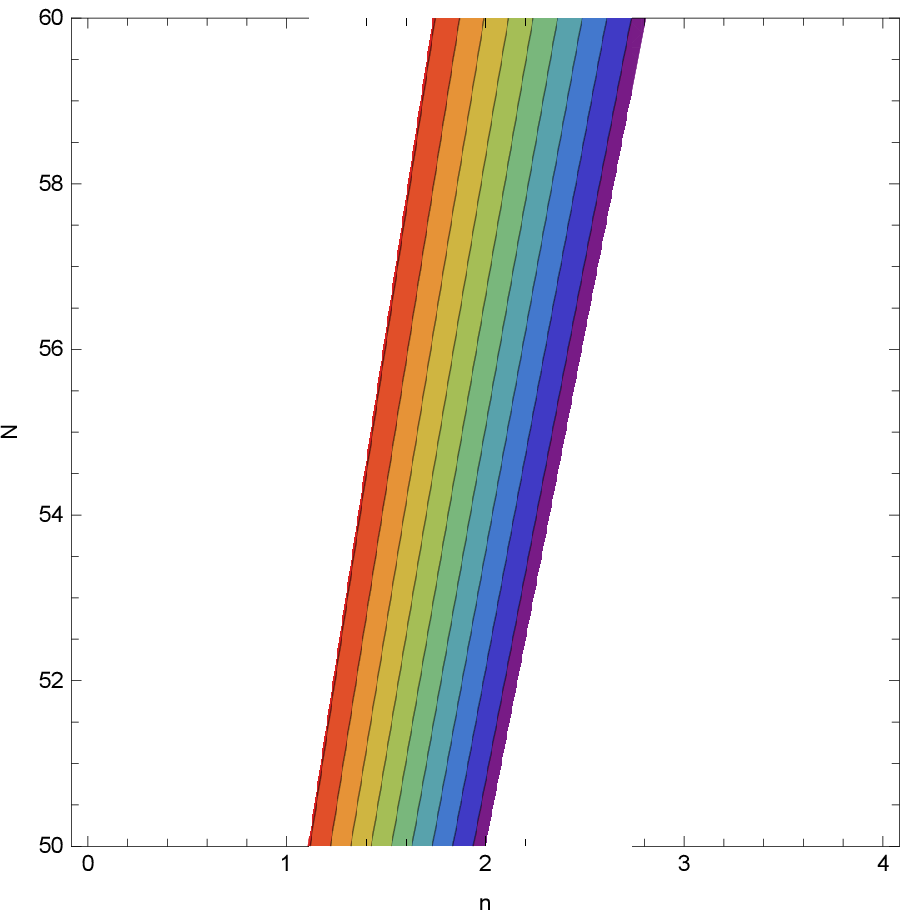}
\includegraphics[width=3pc]{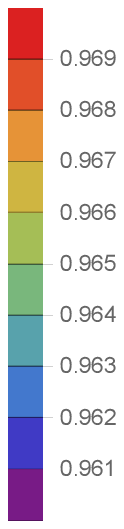}
\caption{Scalar spectral index as a function of the rescale parameter $\alpha$ and the power-law exponent $n$. As shown, compatibility is achieved around the integer value of $n=2$ for the exponent of the scalar field assuming that the e-foldings number resides in the area [50,60].}
\label{scalarACS}
\end{figure}
\begin{equation}
\label{e1}
\centering
\epsilon_1 \simeq \frac{\alpha  n^2}{2 \kappa ^2 \phi ^2}  ,
\end{equation}
\begin{equation}
\label{e2}
\centering
\epsilon_2\simeq -\frac{\alpha  (n-2) n}{2 \kappa ^2 \phi ^2},
\end{equation}

\begin{equation}
\label{e3}
\centering
\epsilon_3\simeq -\frac{2 \alpha  \kappa ^2  n^3 p^2 V_2^2 (n+p-2) (\kappa  \phi )^{2 (n+p)}}{4 \kappa ^4  n^2 p^2 \phi ^2 V_2^2 (\kappa  \phi )^{2 (n+p)}-9 \alpha ^2 \phi ^6} .
\end{equation}
\begin{figure}[h!]
\centering
\includegraphics[width=15pc]{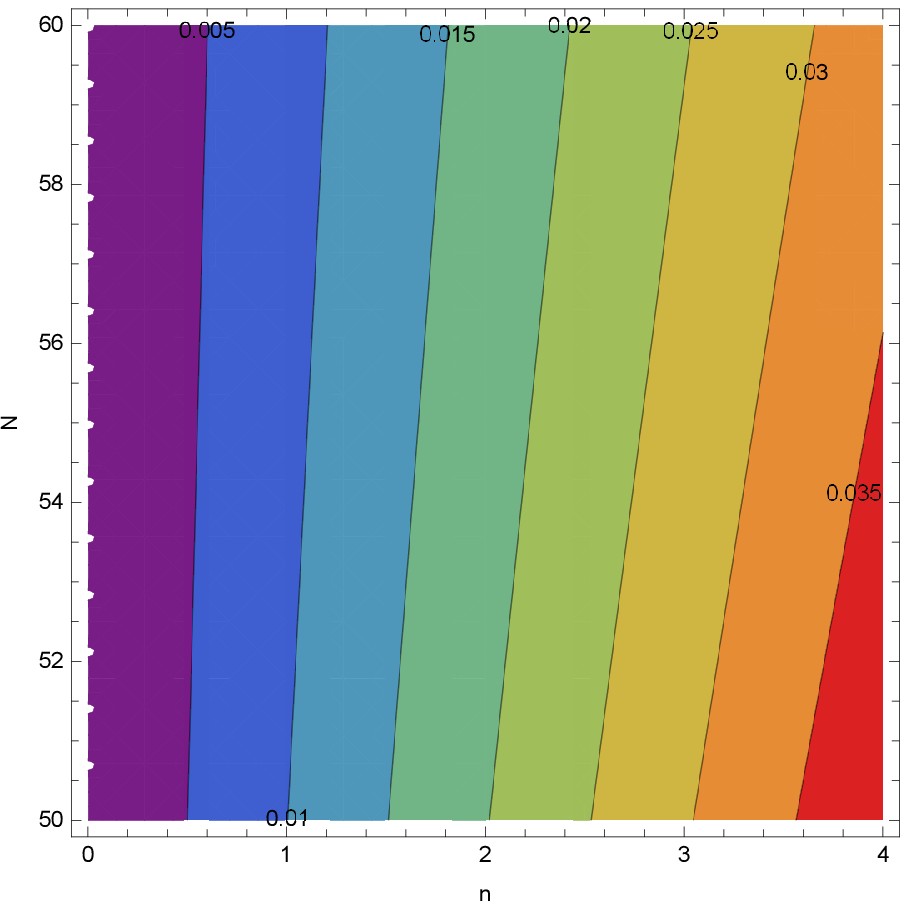}
\includegraphics[width=15pc]{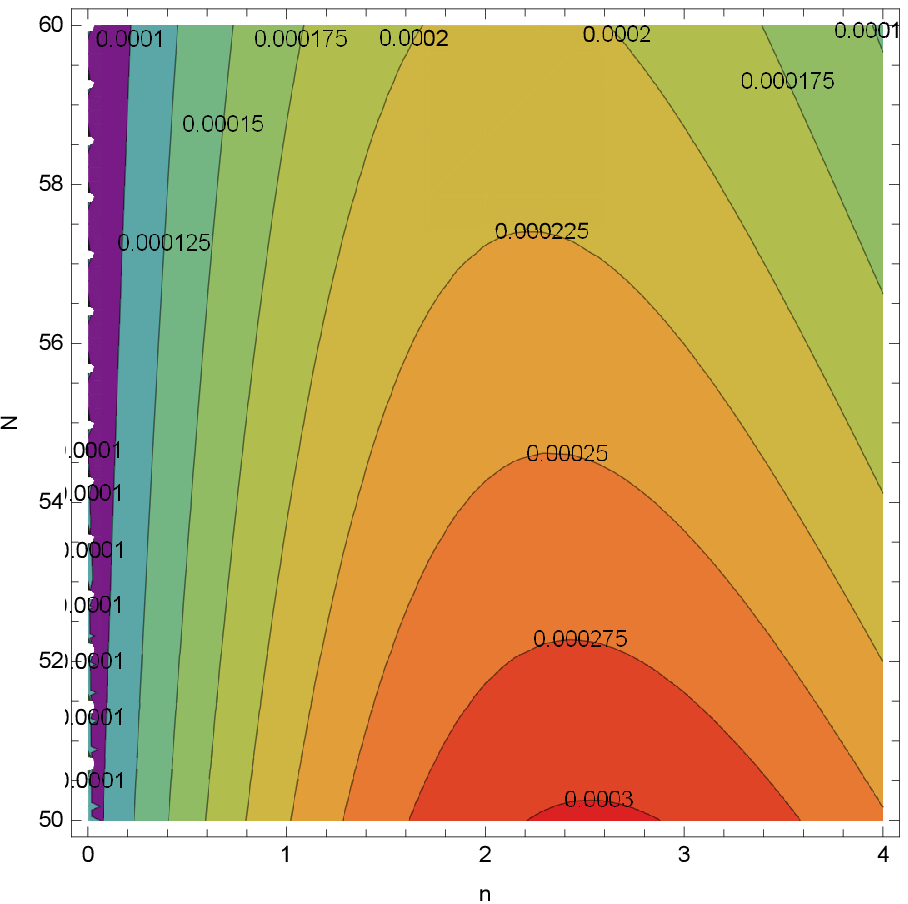}
\caption{Tensor spectral index (left) for $p=4$ and tensor to scalar ratio (right) for $p=2$ as functions of the exponent $n$ and the e-foldings number $N$. It becomes clear that the tensor spectral index obtains a positive value for a plethora of values of the free parameters}
\label{tensorACS1}
\end{figure}
As it is obvious from the above equations, the first two slow-roll parameters are given by quite simple expressions while, $\epsilon_3$ seems bizarre due to the involvement of the parity violating term. Also, as expected, the first two slow-roll indices are independent of $p$, given that the Chern-Simons scalar coupling function is absent from the equations of motion. Now, in order to determine the final value of the scalar field at the end of the inflationary era, we impose the condition $\epsilon_1=1$ hence,
\begin{equation}
\centering
\label{phifinalpowerlaw}
\phi_f=\frac{n}{\kappa}\sqrt{\frac{\alpha}{2}}.
\end{equation}
Utilizing Eq. (\ref{efolds2}), the initial value of the scalar field during the first horizon crossing is
\begin{equation}
\centering
\label{phinitialpowerlaw}
\phi_i=\frac{\sqrt{\alpha  n (n+4 N)}}{\sqrt{2} \kappa}.
\end{equation}
\begin{figure}[h!]
\centering
\includegraphics[width=15pc]{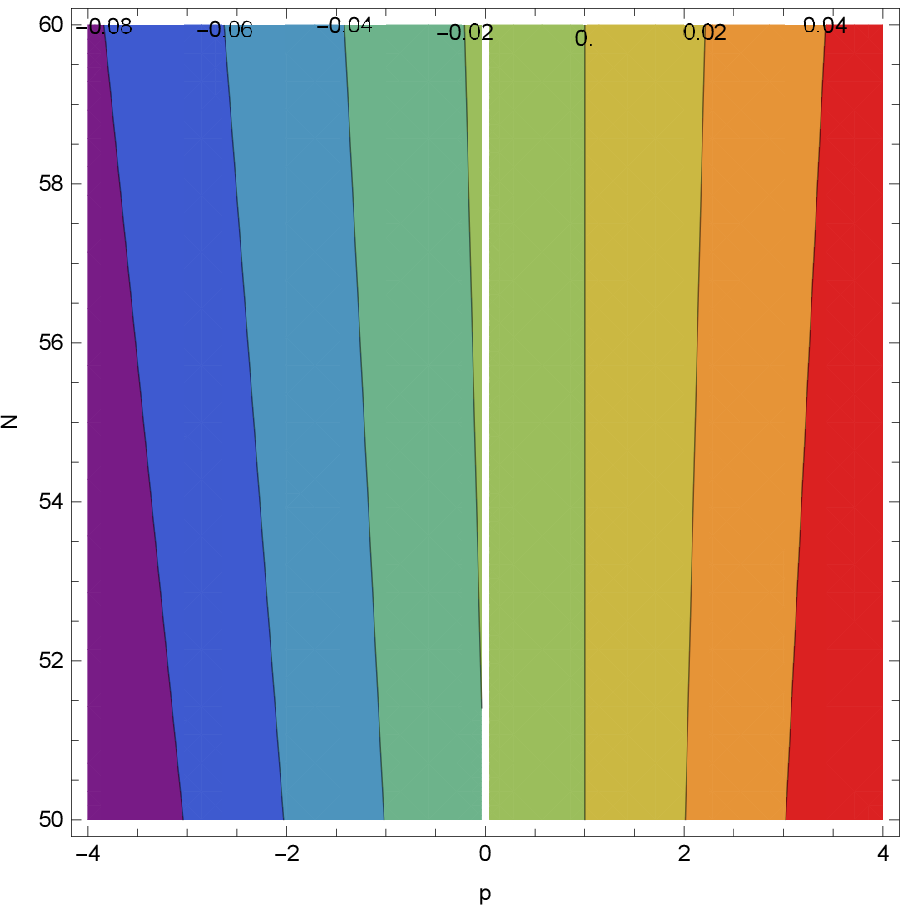}
\includegraphics[width=15pc]{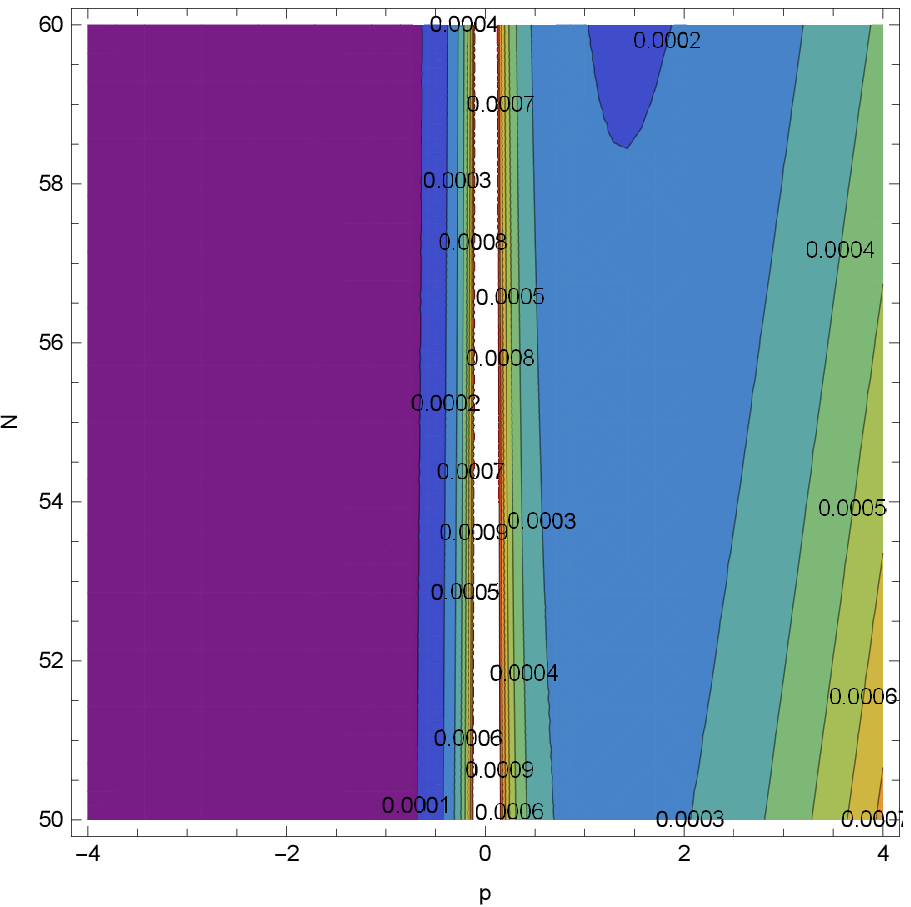}
\caption{Tensor spectral index $n_\mathcal{T}$ and tensor to scalar ratio $r$ as functions of the Chern-Simons exponent $p$ and e-folds. As expected, the Chern-Simons scalar coupling functions has a major impact on the aforementioned observed indices.}
\label{tensorACS2}
\end{figure}
At this point of our analysis, the free parameters of the theory are being specified in order to showcase compatibility with the observational constraints. We consider the following set of numerical values $(N, p, n, V_2, \alpha)=\left(60, 2, 2, M_P^4, 0.001\right)$ for the free parameters in reduced Planck units, where $\kappa^2=1$. The exponents are chosen so that the model is symmetric in the change $\phi\to-\phi$ for simplicity. In this case, the sign between the initial and final values of the scalar field is not relevant but is chosen in such a way that it decreases as time flows by. According to the aforementioned numerical values, the model can be considered as viable, given that the scalar spectral index of primordial perturbations, the tensor spectral index and the tensor to scalar ratio take the values $n_\mathcal{S}=0.966942$, $n_\mathcal{T}=4\cdot 10^{-8}$ and $r=0.000205$, respectively. In Fig.\ref{scalarACS} the scalar spectral index is depicted as a function of two free parameters; in particular, the exponent $n$ and the e-foldings number $N$.

\begin{figure}[h!]
\centering
\includegraphics[width=15pc]{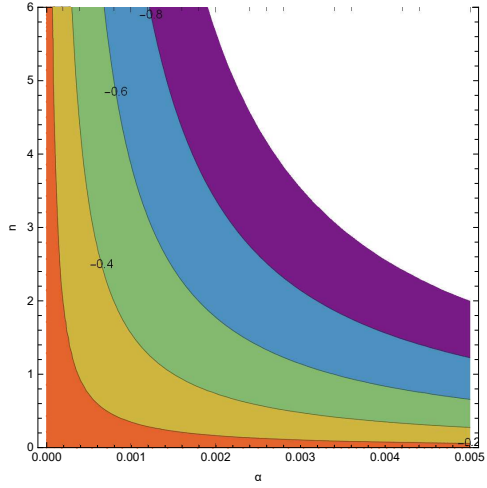}
\caption{Diagrammatic representation of the difference $\kappa\Delta\phi$ with respect to $\alpha$ and $n$. In this case, the conjecture $|\kappa\Delta\phi|<\mathcal{O}(1)$ is satisfied either for large $\alpha$ and small $n$ or the opposite. Compatibility with the Planck 2018 collaboration data limits the options to $n\sim2$ and $\alpha$ being quite small.}
\label{swampACS1}
\end{figure}

\begin{figure}[h!]
\centering
\includegraphics[width=15pc]{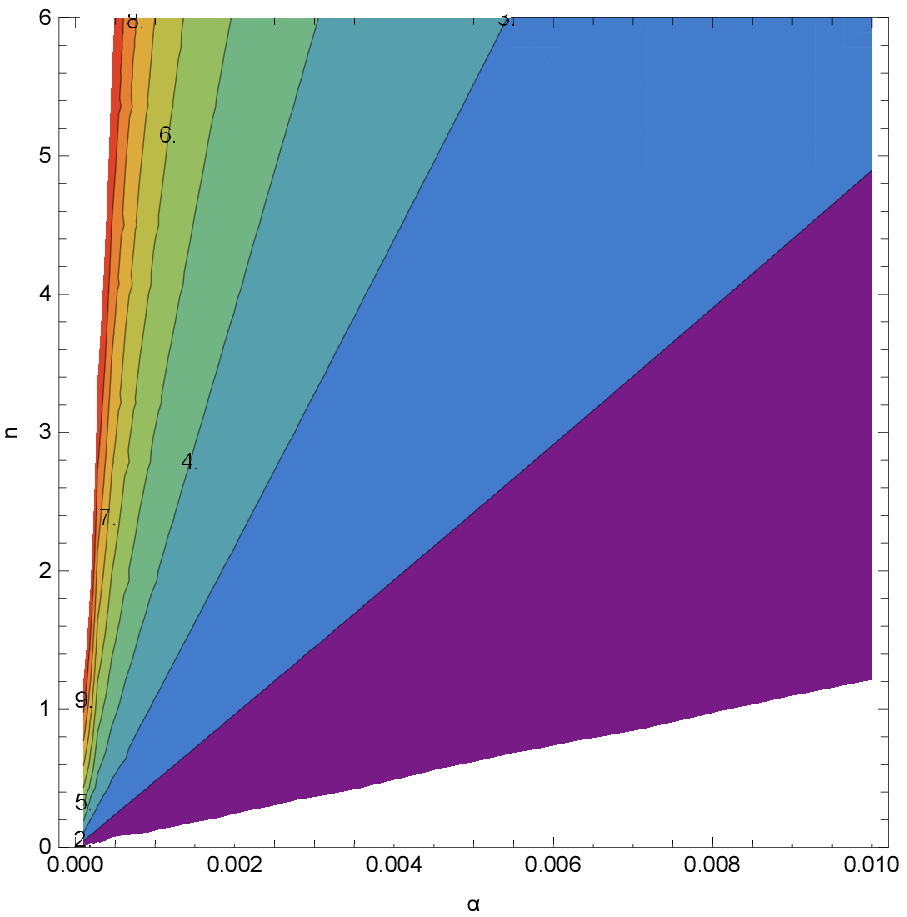}
\includegraphics[width=15pc]{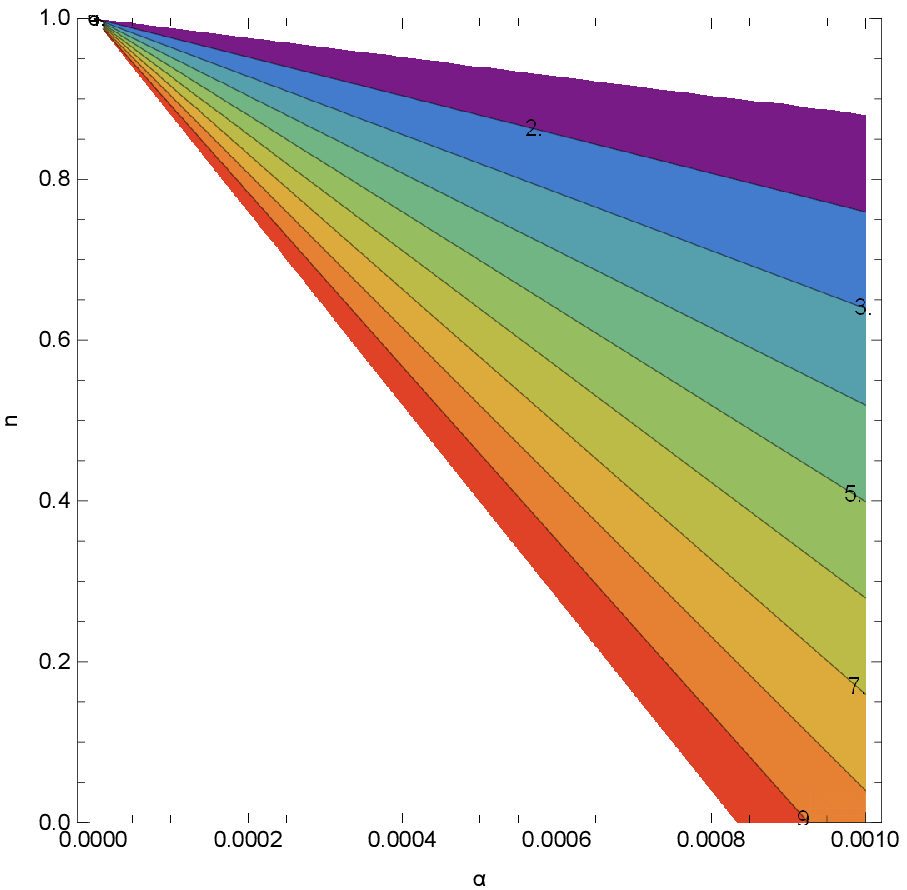}
\caption{Contour plots of $\frac{V'(\phi_i)}{\kappa V{\phi_i}}$ (left) and $-\frac{V''(\phi_i)}{\kappa^2 V(\phi_i)}$ (right) with respect to $\alpha$ and $n$. It becomes clear that no pair of values can be deducted such that both conjectures are satisfied along with the condition $|\kappa\Delta\phi|<\mathcal{O}(1)$ and at the same time be in agreement with the Planck 2018 collaboration data, since the third condition requires an exponent $0<n<1$.}
\label{swampACS2}
\end{figure}

It is interesting to comment on the scalar spectral index, since it is essentially independent of the rescale parameter $\alpha$. This property was also observed in the canonical scalar field in Ref. \cite{Oikonomou:2021zfl} and was expected in the Chern-Simons case as well, since only tensor perturbations are affected. In addition, Fig.\ref{tensorACS1} and Fig.\ref{tensorACS2} depict the tensor spectral index and the tensor to scalar ratio. It is worth mentioning that the model at hand predicts a blue tilted tensor spectral index for a variety of pairs of values, both for the Chern-Simons exponent $p$ which has a major impact on these observables and also the exponent of the scalar field $n$. In contrast, the e-folding number has a mild influence on the tensor spectral index. Moreover, the initial and the final numerical values of the scalar field are $\kappa\phi_i=0.491935$ and $\kappa\phi_f=0.0447214$ respectively. Hence, for the case at hand, one can easily see that the condition $|\kappa\Delta\phi|<\mathcal{O}(1)$ is satisfied for the aforementioned parameters and in Fig.\ref{swampACS1} the diagrammatic representation of said difference is pointed out with respect to $n$ and $\alpha$. Furthermore, the rest of the conditions are not satisfied for the exact same parameters since $\frac{V'(\phi_i)}{\kappa V(\phi_i)}=4.06558$ which is indeed satisfied however, $-\frac{V''(\phi_i)}{\kappa^2V(\phi_i)}=-8.26446$ which is not an acceptable value. In general, these two conditions are opposite and cannot be satisfied together. This becomes clear in Fig.\ref{swampACS2}, where the previously mentioned conditions are illustrated. In general, if even a single condition is satisfied then the Swampland criteria are also satisfied. Therefore, the inclusion of a Chern-Simons term in the choice of a power-law scalar potential not only does it make the model compatible with the Planck data (given that the tensor to scalar ratio transitions below the threshold $0.064$ for a viable scalar spectral index) but is also compatible with the Swampland criteria, regardless of all three conjectures not being satisfied simultaneously. In addition, a blue tilted tensor spectral index may be generated as indicated previously, but it is not mandatory. It should be stated that since the theory is symmetric in $\phi\to-\phi$, using the negative value for $\phi_i$ affects only the first two conditions by changing their sign. However, the third is not affected as one would expect since $V''$ is independent of $\phi$ and $V$ is symmetric in $\phi$ for $n=2$. Note also that the third condition, which is not satisfied for the parameters used here, is actually at variance with the viability of the scalar spectral index, since the first is respected for exponents below unity while the latter for exponents close to $n=2$.

Finally, we mention that the slow-roll indices obtain the values $\epsilon_1\sim \mathcal{O}\left(10^{-3}\right)$, $\epsilon_2 =0$ and $\epsilon_3 \sim \mathcal{O}\left(10^{-3}\right)$ thus justifying the slow-roll assumption.

\subsection{Axionic inflation with a Power-law Chern-Simons scalar coupling function}

The second model we study in the context of Einstein-Chern-Simons gravity involves a scalar coupling function of the form
\begin{equation}
\label{couplingaxionic}
\nu(\phi)=\Lambda(\kappa \phi)^q,
\end{equation}
where in contrast to the previous model, the prefactor is considered to be present in order to observe the impact on the tensor to scalar ratio. For simplicity, $\Lambda$ is a dimensionless parameter. The scalar potential has the axionic-inspired form according to Ref. \cite{Pajer:2013fsa}
\begin{equation}
\label{axionicpotential}
\centering
V(\phi)=V_3\bigg[1+\cos\bigg(\frac{\phi}{\phi_0}\bigg)\bigg],
\end{equation}
\begin{figure}[h!]
\centering
\includegraphics[width=15pc]{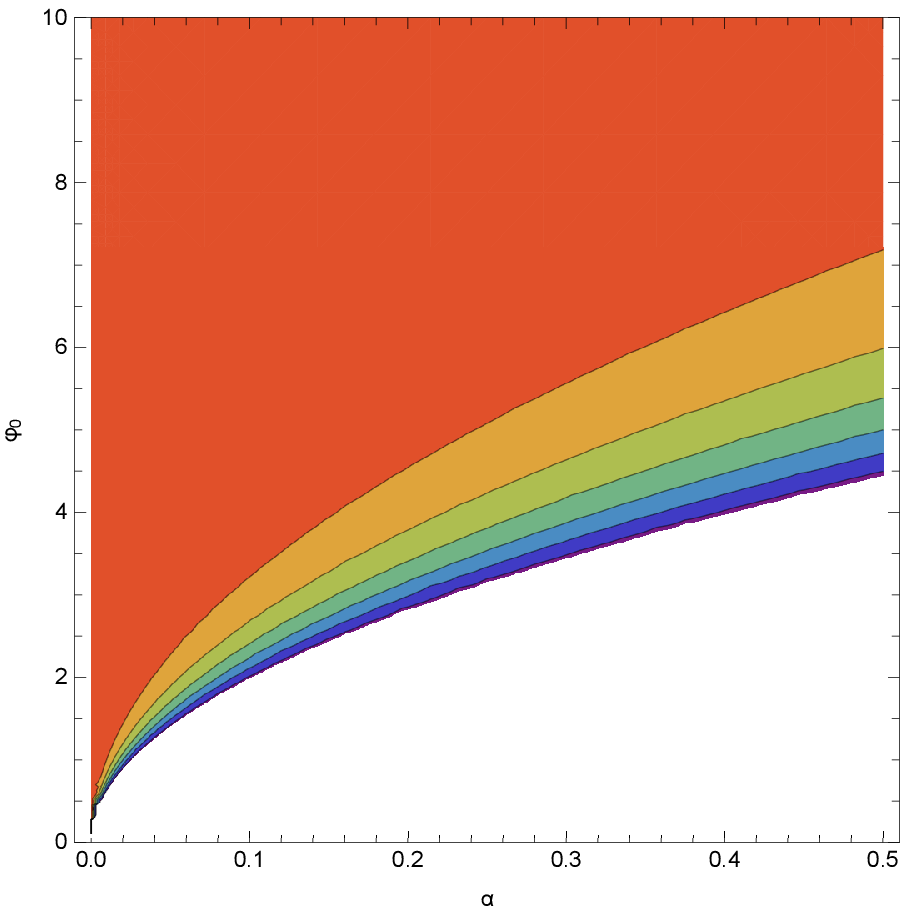}
\includegraphics[width=4pc]{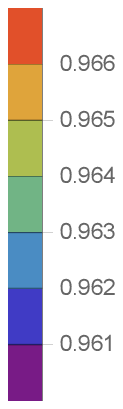}
\includegraphics[width=15pc]{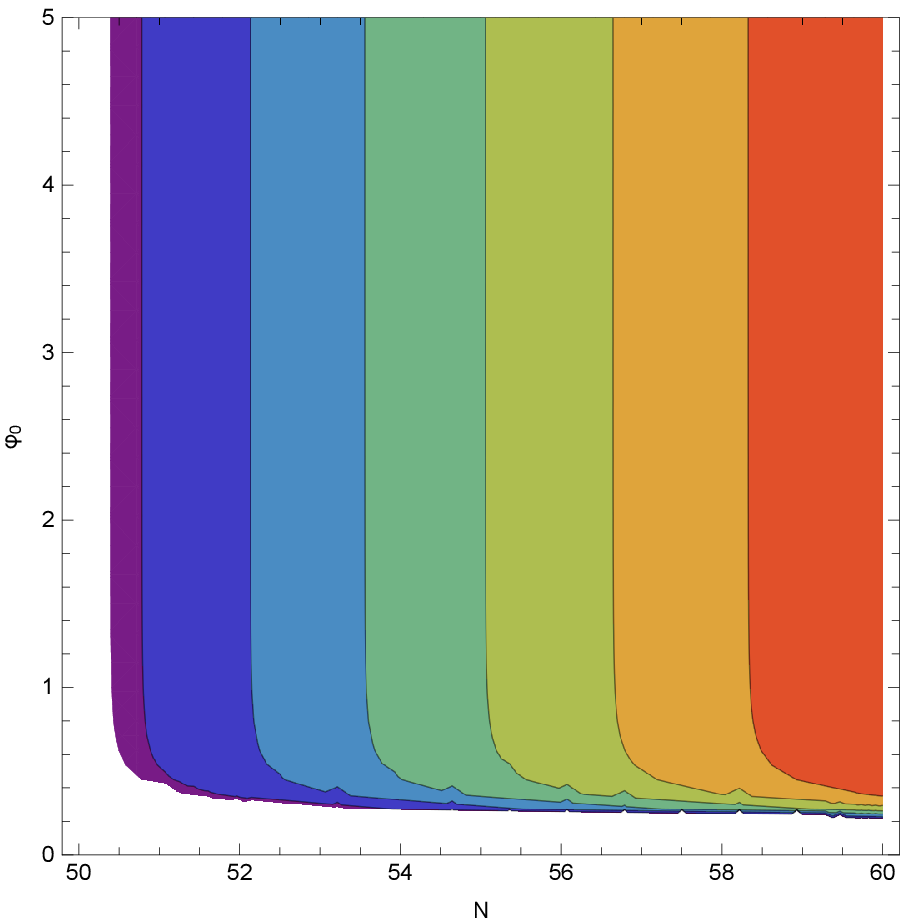}
\includegraphics[width=4pc]{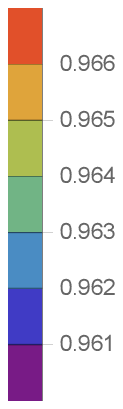}
\caption{Dependence of the scalar spectral index $n_\mathcal{S}$ on the parameters $\alpha$, $\phi_0$ on the left and $N$, $\phi_0$ on the right. As showcased, there are several values of the free parameters that render the model compatible with the observations.}
\label{scalarBCS}
\end{figure}
where $\phi_0$ is a parameter of the model with dimensions $[\phi_0]=\ev$ and $[V_3]=\ev^4$ represents the amplitude of the scalar potential. For simplicity, the amplitude of axionic potential shall be considered to be equal to unity in natural units.

The slow-roll indices during the inflationary era are given by the expressions
\begin{equation}
\label{epsilon1axionic}
\centering
\epsilon_1 \simeq \frac{\alpha  \tan ^2\left(\frac{\phi }{2 \phi_0}\right)}{2 \kappa ^2 \phi_0^2},
\end{equation}
\begin{equation}
\label{epsilon2axionic}
\centering
\epsilon_2 \simeq \frac{\alpha }{2 \kappa ^2 \phi_0^2},
\end{equation}

\begin{equation}
\label{epsilon3axionic}
\centering
\epsilon_3 \simeq \frac{4 \alpha  \kappa ^2 q^2 (\kappa  \phi )^{2q} \sin ^2\left(\frac{\phi }{2 \phi_0}\right) \left((q-1) \phi_0 \sin \left(\frac{\phi }{\phi_0}\right)+\phi  \cos \left(\frac{\phi }{\phi_0}\right)\right)}{4 \kappa ^4 q^2 \phi  \phi_0^2 (\kappa  \phi )^{2 q} \sin ^2\left(\frac{\phi }{\phi_0}\right)-9 \alpha ^2 \phi ^3 \phi_0^4}, 
\end{equation}
where the complicated form of index $\epsilon_3$ is due to the inclusion of the parity violating term. Utilizing the formalism which in the previous models, the initial and the final values of the inflaton are, respectively,
\begin{figure}[h!]
\centering
\includegraphics[width=15pc]{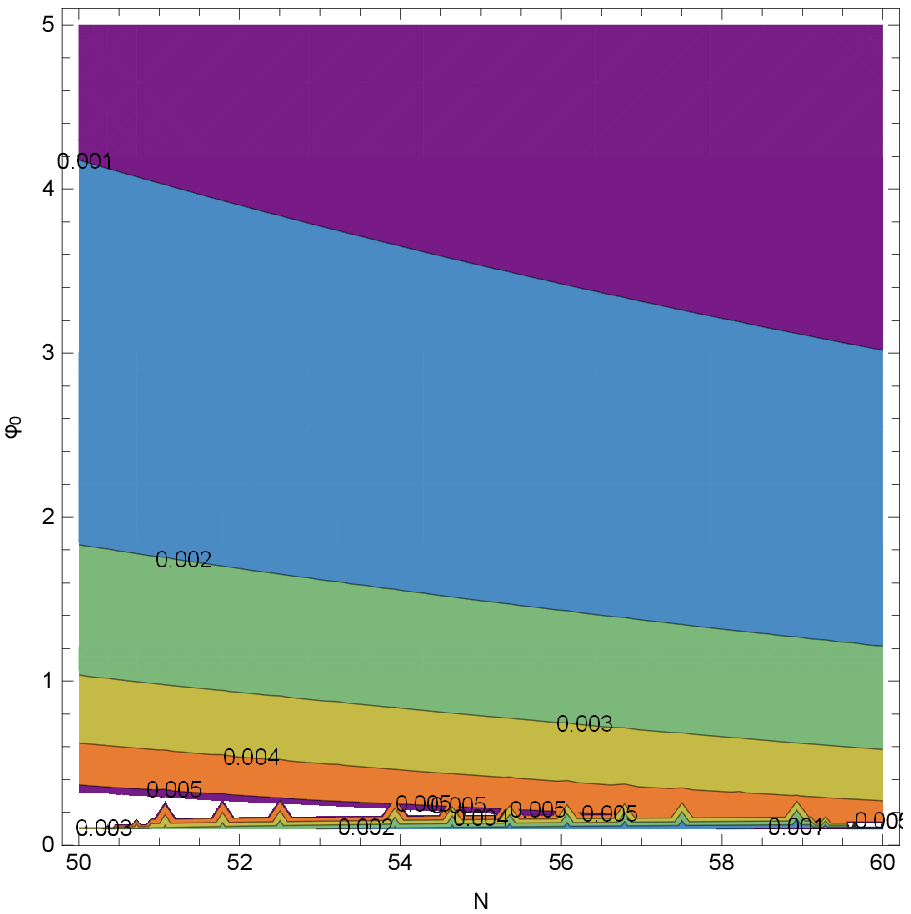}
\includegraphics[width=15pc]{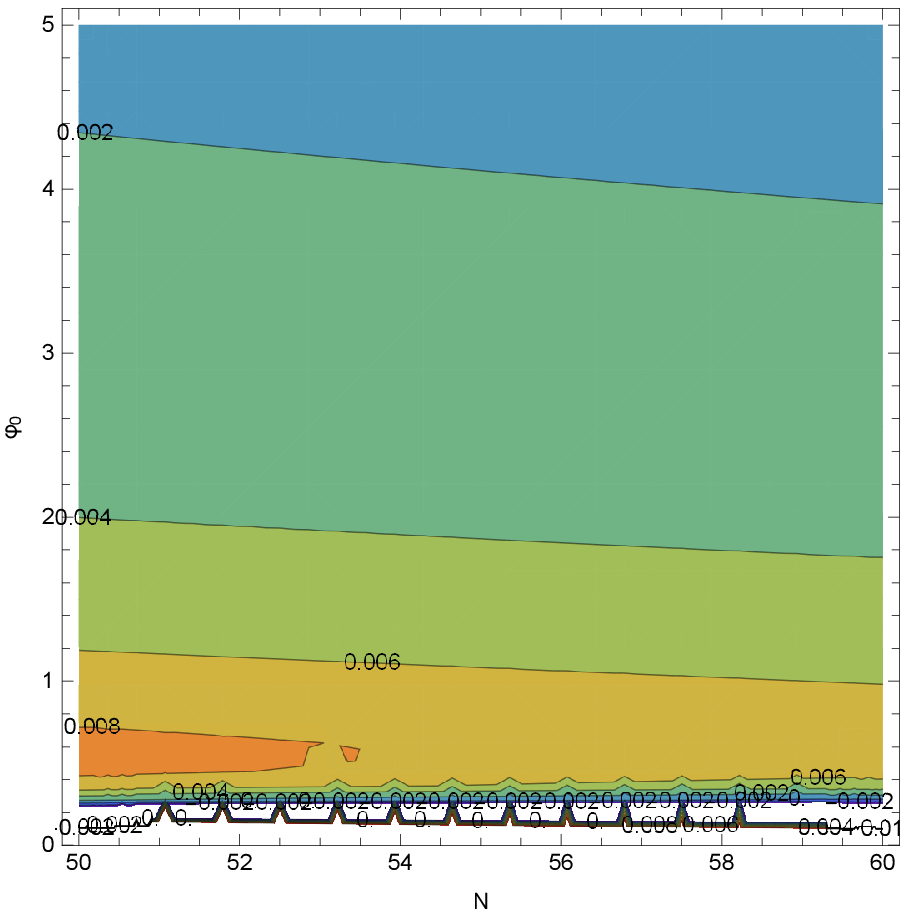}
\caption{Tensor to scalar ratio $r$ (left) and tensor spectral index $n_\mathcal{T}$ as functions of parameter $\phi_0$ and the e-foldings number $N$. If $\Lambda$ were to be increased then the tensor to scalar ratio decreases in turn. }
\label{tensorBCS}
\end{figure}
\begin{equation}
\centering
\label{phiinitialaxionic}
\phi_i=2 \phi_0 \sin ^{-1}\left\lbrace e^{-\frac{\alpha  N}{2 \kappa ^2 \phi_0^2}} \sin \left[\frac{1}{2} \tan ^{-1}\left(\frac{2 \alpha }{\alpha +2 \kappa ^2 \phi_0^2}-1,\frac{2 \sqrt{2} \sqrt{\alpha } \kappa  \phi_0}{\sqrt{\left(\alpha +2 \kappa ^2 \phi_0^2\right)^2}}\right)\right]\right\rbrace,
\end{equation}

\begin{equation}
\centering
\label{phifinalaxionic}
\phi_f=\phi_0 \tan ^{-1}\left(\frac{2 \alpha }{\alpha +2 \kappa ^2 \phi_0^2}-1,\frac{2 \sqrt{2} \sqrt{\alpha } \kappa  \phi_0}{\sqrt{\left(\alpha +2 \kappa ^2 \phi_0^2\right)^2}}\right).
\end{equation}

\begin{figure}[h!]
\centering
\includegraphics[width=15pc]{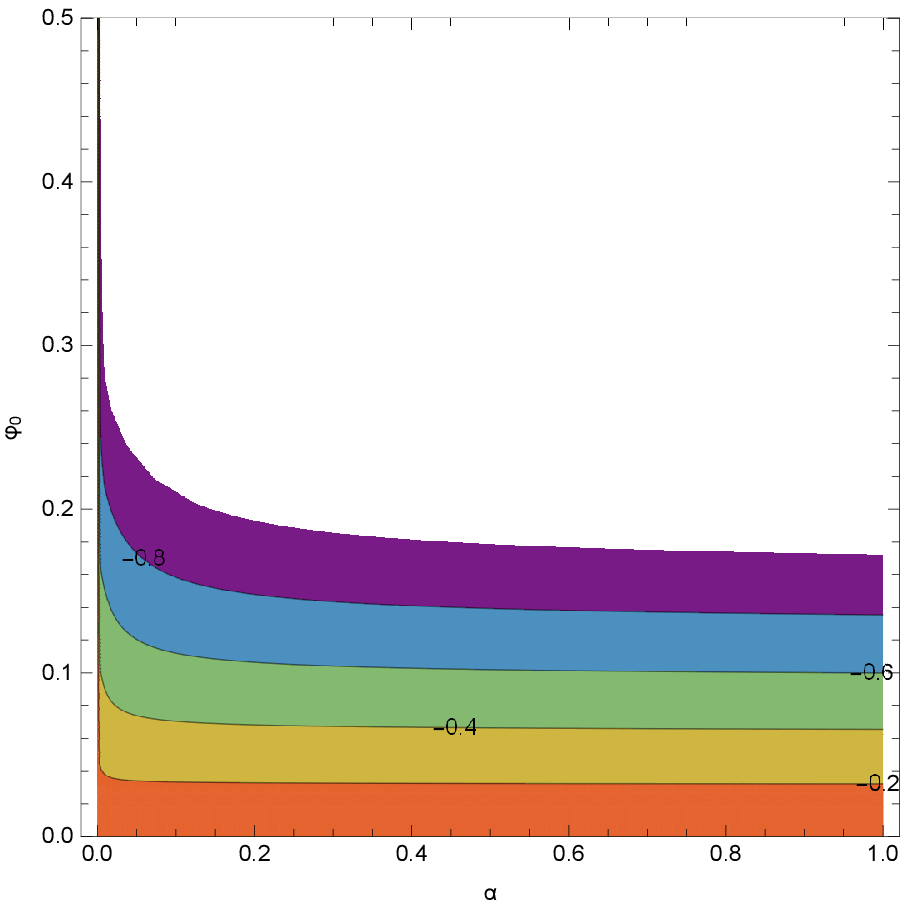}
\includegraphics[width=15pc]{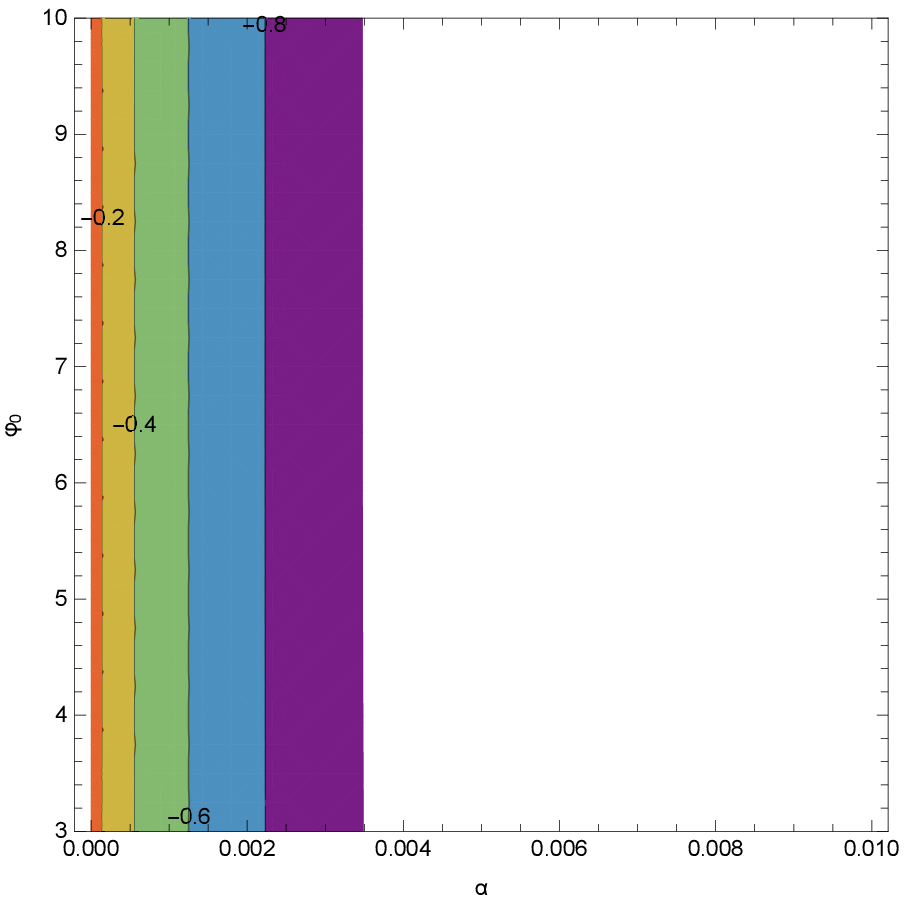}
\caption{Scalar field difference $\kappa\Delta\phi$ as a function of $\alpha$ and $\phi_0$. It becomes clear that this Swampland criterion is satisfied either for small $\phi_0$ and large $\alpha$ (left) or vice-versa (right), which is indeed in agreement with the Planck data.}
\label{crit1BCS}
\end{figure}
Let us now proceed to the designation of the free parameters in order to ascertain whether the model is compatible with the Planck data, while it satisfies the Swampland criteria at the same time. By imposing the values $(N, \phi_0, \alpha, \Lambda, V_3, q)=\left(60, 12 M_P, 0.0012, 0.001, M_P^4, 4\right)$ then the scalar spectral index obtains the value $n_\mathcal{S}=0.966942$, while the tensor spectral index and the tensor to scalar ratio are now equal to $n_\mathcal{T}=-0.00791792$ and $r=0.00028381$ respectively, which are indeed in agreement with \cite{Planck:2018vyg}. The free parameter $q$ needs to be quite large since $\phi_0$ is likewise large; however, a different set of parameters, as we shall present later, is also a feasible scenario. In Fig.\ref{scalarBCS}, the scalar spectral index depending on $N$, $\phi_0$ and $\alpha$ is depicted, from which it can easily be inferred that compatibility can be achieved relatively easy in the range of $N\sim50-60$, for a wide range of values for $\phi_0$ and $\alpha$. In addition, a red tilted tensor spectral index is now obtained (see Fig.\ref{tensorBCS}) only because the relation $\epsilon_3<-\epsilon_1$ during the first horizon crossing does not hold true. It should be stated that a larger exponent $q$ decreases the tensor to scalar ratio while it can simultaneously generate a blue tilted tensor spectral index. In Fig.\ref{tensorBCS}, a larger $\Lambda$ is considered in order to illustrate the impact of said parameter on the tensor spectral index where, as shown, this particular increase results in the decrease of the tensor to scalar ratio. This is indicative of the impact of the Chern-Simons scalar coupling function on the tensor to scalar ratio.

At this stage, it is interesting to comment on the numerical value of the Swampland conjectures. For the aforementioned parameters, one can easily check that $\kappa\phi_i=38.2367$ and $\kappa\phi_f=37.6502$ which suggests that $|\kappa\Delta\phi|<1$ and thus this criterion is satisfied. The value of the scalar field during the first horizon crossing and at the final stage of the inflationary era is of order $\mathcal{O}(10)$, since a large value of $\phi_0$ was considered. One can easily satisfy both the Swampland criterion $|\kappa\Delta\phi|<\mathcal{O}(1)$ and the Planck data for smaller values of $\phi_0$, as shown in Fig.\ref{crit1BCS}. Furthermore, the numerical value of the rest conditions for the same set of free parameters as before are $\frac{V'(\phi_i)}{\kappa V(\phi_i)}=3.71942$ and $-\frac{V''(\phi_i)}{\kappa^2V(\phi_i)}=-6.91356$. Therefore, as was the case with the power-law model, not all conditions are simultaneously met for the same set of free parameters. However, it should be noted that the third condition $-\frac{V''(\phi_i)}{\kappa^2V(\phi_i)}>\mathcal{O}(1)$ is satisfied if one decreases the value of $\phi_0$. As an example, for $\phi_0=0.24$ in natural units, $-\frac{V''(\phi_i)}{\kappa^2V(\phi_i)}=5.21718$. As a result, one can always find values that are acceptable in order to satisfy all conjectures at the same time. For these numerical values, the tensor spectral index is expected to still be red tilted as now $n_\mathcal{T}=-0.00687946$, but the condition $r=-8n_\mathcal{T}$ is not met due to the Chern-Simons term. In return, the tensor to scalar ratio is decreased to $r=0.00417417$. Note also that a decreasing $q$ results in an increase in $r$. For instance, having $q=1$ generates $r=0.0407654$ and $n_\mathcal{T}=-0.0163336$. It is worth mentioning that both the Planck data and all three conditions for the Swampland criteria can be complied, while a blue tilted tensor spectral index is still expected, just by increasing the exponent of the Chern-Simons scalar coupling function. As an example, having $\phi_0=0.24M_P$ and a Chern-Simons exponent $q=8$, we find that $n_\mathcal{T}=0.00327839$. This is characteristic of the fact that both a positive and a negative value for the tensor spectral index is a possible outcome however, having such a large exponent seems quite extravagant. The value $\phi_0=12M_P$ that renders the model compatible with Planck data and respects two of the conjectures from the Swampland criteria was initially considered, since it seems to also be in agreement with the results of \cite{Pajer:2013fsa}, where for $N=60$, $\phi_0$ being greater than $10$ for such a scalar potential resides in a $95\%$ CL. Hence, in general one can consider the value $\phi_0=12M_P$, regardless of \ref{criterion3} not holding true, since the Swampland criteria are satisfied even if one of them is respected.

\begin{figure}[h!]
\centering
\includegraphics[width=15pc]{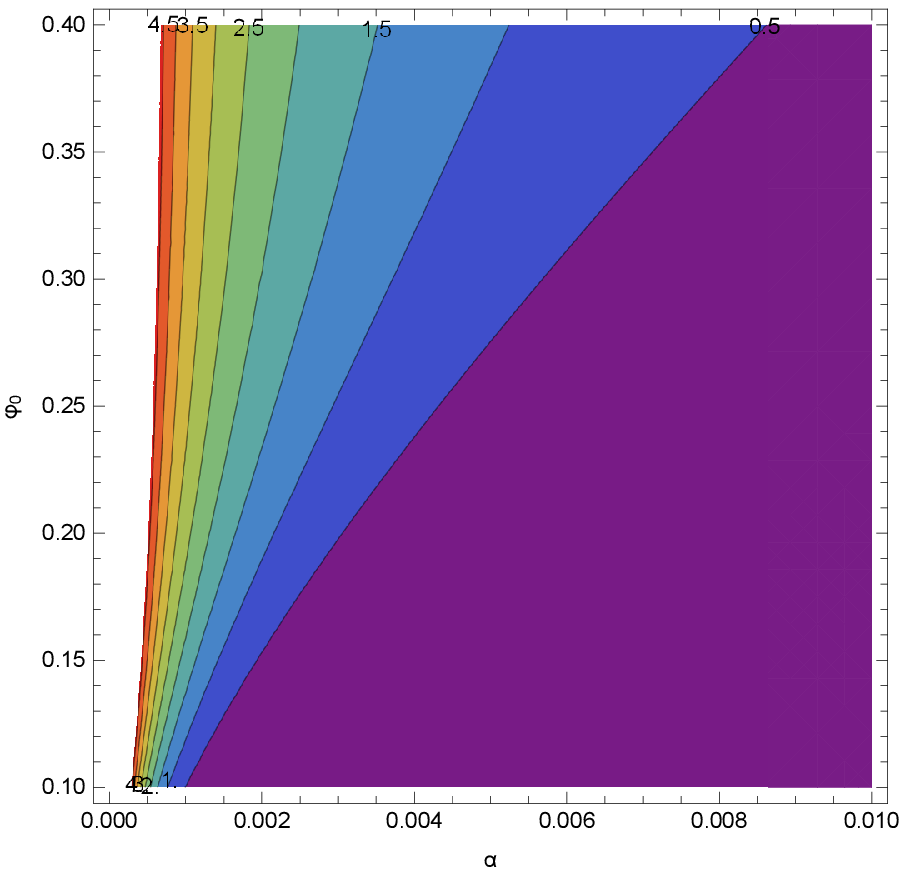}
\includegraphics[width=15pc]{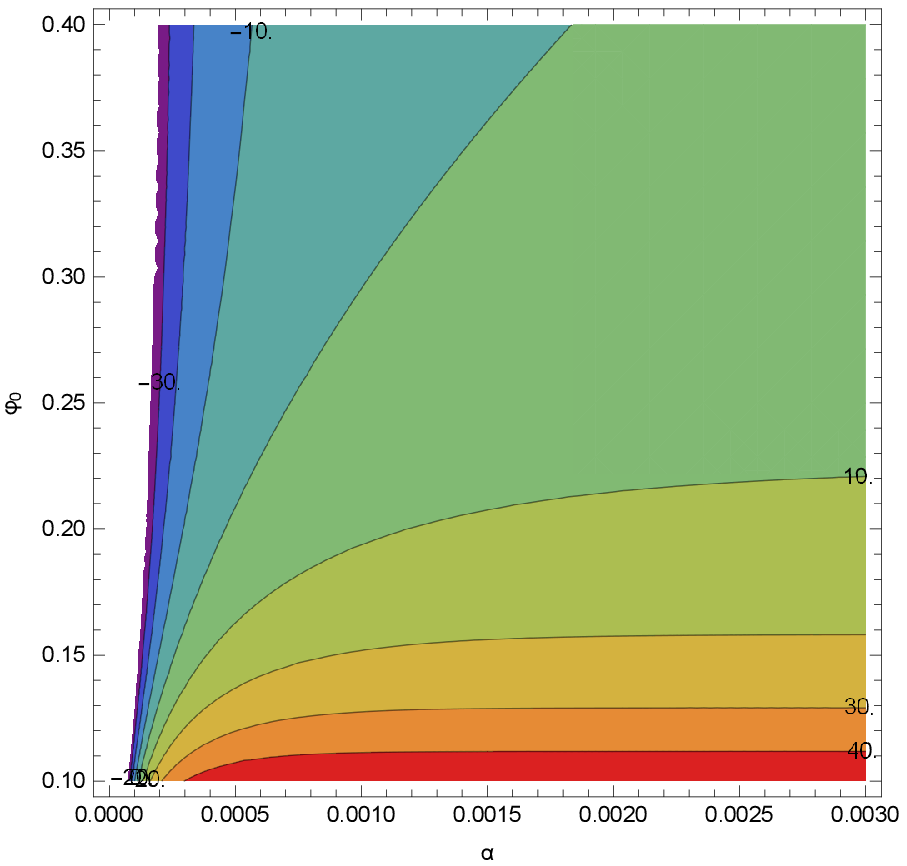}
\caption{Contour plots of the ratios $\frac{V'(\phi_i)}{\kappa V(\phi_i)}$ on the left and $-\frac{V''(\phi)}{\kappa^2V(\phi_i)}$ on the right. As shown, both seem to agree for small values of $\phi_0$ and $\alpha$.}
\label{crit2BCS}
\end{figure}
For consistency, we mention that the slow-roll conditions are satisfied, too. Specifically, the numerical value of the slow-roll indices $\epsilon_1$ and $\epsilon_3$ is of order $\mathcal{O}\left(10^{-3}\right)$, while $\epsilon_2\sim\mathcal{O}\left(10^{-6}\right)$. This is because $\dot H\sim\mathcal{O}\left(10^{-3}\right)$ and $H^2\sim\mathcal{O}\left(10^{-1}\right)$, while $\frac{1}{2}\dot\phi^2\sim\mathcal{O}\left(10^{-6}\right)$ and $V\sim\mathcal{O}\left(10^{-3}\right)$ during the first horizon crossing. These were calculated for the case of $\phi_0=12M_P$ and $q=4$. Now, considering $\phi_0=0.24$ and $q=4$, one should obtain $\epsilon_1=0.004137$, $\epsilon_2$ increases to the value of $0.0103689$ and finally $\epsilon_3=-0.000697$. Decreasing of $\phi_0$ to such a value results in an effective increase of $\dot H$, $H^2$, $\frac{1}{2}\dot\phi^2$ and $V$ by four orders of magnitude. Finally, the case of $q=8$ suggests that $\epsilon_3=-0.0057762$ which is not a dramatic change; it just surpasses in absolute value the first slow-roll index, thus resulting in a blue tilted tensor spectral index.

\begin{figure}[h!]
\centering
\includegraphics[width=15pc]{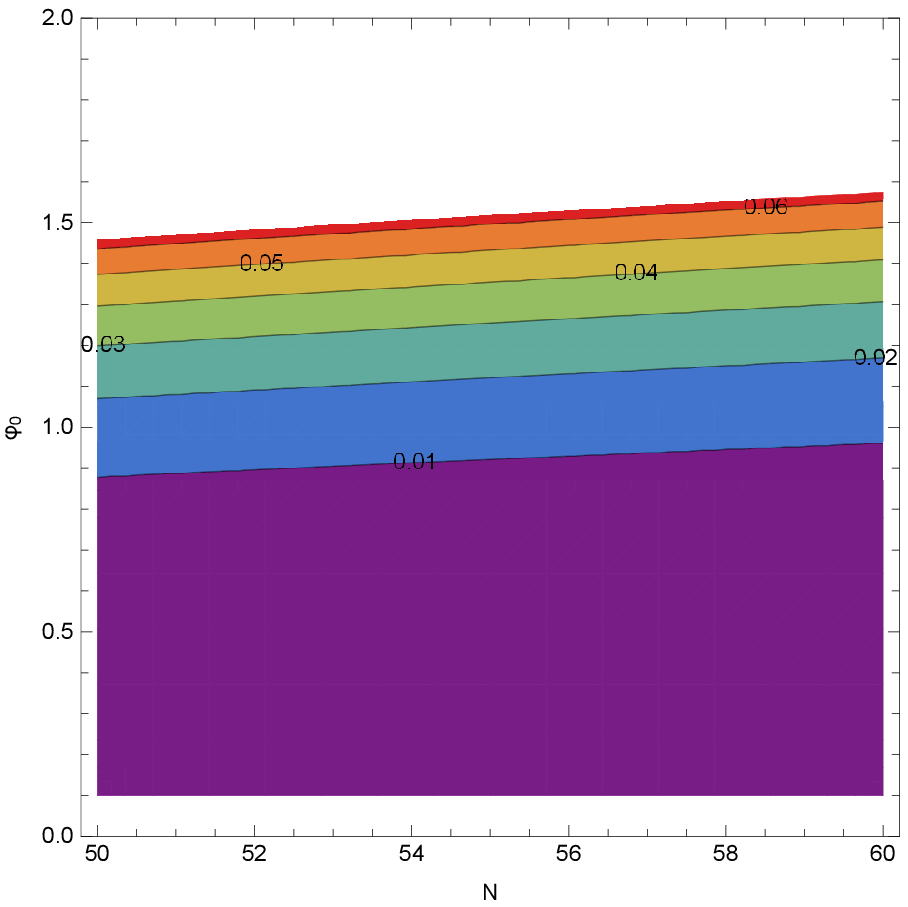}
\includegraphics[width=15pc]{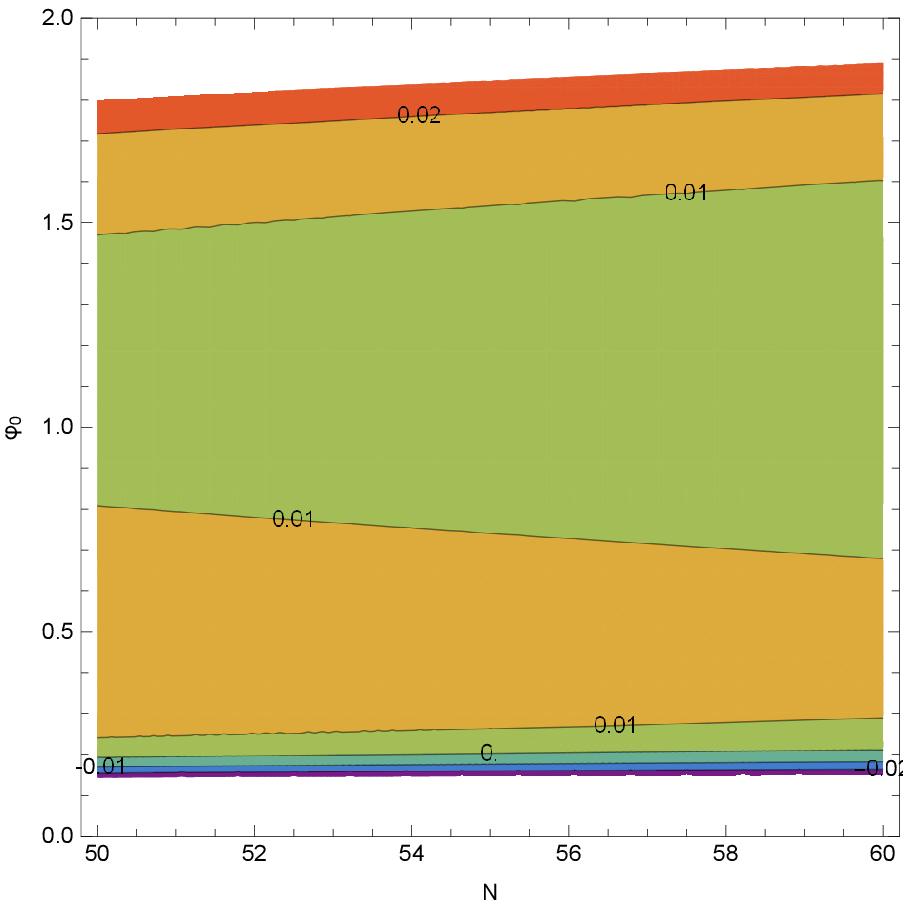}
\caption{Diagrammatic representation of $r$ and $n_\mathcal{T}$ for the case of the exponential Chern-Simons scalar coupling function. The values considered here are $\tilde c=1.14$, $\Lambda=0.001$, $\phi_0=0.24$ with $\alpha$, $V_3$ , $N$ remaining unaffected.}
\label{exp}
\end{figure}

As a final remark, let us discuss the choice of a different Chern-Simons scalar coupling function but with the same set of free parameters as before. Assuming that $\nu(\phi)=\Lambda\exp\bigg(\tilde c\frac{\phi}{\phi_0}\bigg)$, with $\tilde c$ being a dimensionless parameter, the tensor spectral index and the tensor to scalar ratio, which are depicted in Fig.\ref{exp}, are affected in a similar manner as to the case of the power-law coupling. The choice of a different Chern-Simons scalar coupling function, in contrast to the previously presented Gauss-Bonnet case, does not affect the Swampland criteria since it only affects tensor perturbations. In particular, increasing $\Lambda$ seems to effectively decrease the tensor to scalar ratio. As a matter of fact, substituting $\tilde c=1.14$, $\phi_0=0.24M_{P}$ and for either $\Lambda=1$ or $\Lambda=0.001$ one obtains $r=1.85767\cdot10^{-8}$ or $r=0.0000185767$, respectively. No matter the choice of $\Lambda$, the tensor spectral index now reads $n_\mathcal{T}=0.000426986$.

\section{Conclusions}
In the present work, we examined the validity of rescaled $f(R)$ models with string-correction terms. This is a general class that can be used in order to unify early and late time eras, where the scalar field is dominant during the inflationary era while the $f(R)$ part is dominant in the late era. We showed that while it is relatively straightforward to obtain results compatible with the Planck data for a plethora of values for the rescale parameter $\alpha$, which is assumed to reside in the interval $0<\alpha<1$, the Swampland criteria are satisfied only for small values of $\alpha$, both for the case of the constrained Gauss-Bonnet and the Chern-Simons models, which is similar to the case of a canonical scalar field; however, even though the conjecture $|\kappa\Delta\phi|<1$ is satisfied, the rest of the conditions are not necessarily met simultaneously as shown especially in the Chern-Simons case. Due to the string corrections considered, a blue tilted tensor spectral index can be generated for viable models with the Planck data while the Swampland criteria are satisfied. This is an interesting feature that is connected to the amplification of the energy spectrum of primordial tensor perturbations, which may finally be observed in subsequent years. An interesting question that arises is whether the Swampland criteria are satisfied, while the model is in agreement with Planck data, for larger values of $\alpha$, if one considers theories with additional terms such as a non-canonical kinetic term,  or even under the assumption that the scalar field satisfies the constant-roll condition $\ddot\phi=\beta H\dot\phi$. The latter case is also connected to the possible detection of non-Gaussianities in the CMB and we hope to address this in a future project.

\end{document}